\documentclass[lettersize,journal]{IEEEtran}
\usepackage{amsmath,amsfonts}
\usepackage{algorithmic}
\usepackage{algorithm}
\usepackage{array}
\usepackage[caption=false,font=normalsize,labelfont=sf,textfont=sf]{subfig}
\usepackage{tikz}
\usetikzlibrary{arrows.meta}
\usepackage{textcomp}
\usepackage{stfloats}
\usepackage{url}
\usepackage{verbatim}
\usepackage{graphicx}
\newcommand {\Define} {\stackrel {\Delta} {=}  }
\newcommand{\mya}{\mathrel{\overset{\makebox[0pt]{{\tiny(a)}}}{=}}}
\newcommand{\myb}{\mathrel{\overset{\makebox[0pt]{{\tiny(b)}}}{=}}}

\newtheorem{theorem}{Theorem}

\newtheorem{lemma}{Lemma}

\usepackage{cite}
\usepackage{graphicx}
\usepackage{caption}
\usepackage{subcaption}
\usepackage{lipsum} 
\usepackage{tabularx}
\usepackage{subfig}
\usetikzlibrary{arrows.meta}
\begin{document}

\title{Zak-OTFS for Identification of Linear Time-Varying Systems}

\author{Danish Nisar,~\IEEEmembership{Staff,~IEEE,}
\thanks{S. K. Mohammed is with the }
}

\author{\IEEEauthorblockN{Danish Nisar\IEEEauthorrefmark{1}, Saif Khan Mohammed\IEEEauthorrefmark{1},
Ronny Hadani\IEEEauthorrefmark{2}\IEEEauthorrefmark{3},
Ananthanarayanan Chockalingam\IEEEauthorrefmark{4} and 
Robert Calderbank\IEEEauthorrefmark{5}~\IEEEmembership{Fellow,~IEEE}}
\\
\IEEEauthorblockA{\IEEEauthorrefmark{1}Department of Electrical Engineering, Indian Institute of Technology Delhi, India}\\
\IEEEauthorblockA{\IEEEauthorrefmark{2}Department of Mathematics, University of Texas at Austin, USA}\\
\IEEEauthorblockA{\IEEEauthorrefmark{3}Cohere Technologies Inc., CA, USA}\\
\IEEEauthorblockA{\IEEEauthorrefmark{4} Department of Electrical and Communication Engineering, Indian Institute of Science Bangalore, India}\\
\IEEEauthorblockA{\IEEEauthorrefmark{5}Department of Electrical and Computer Engineering, Duke University, USA}\\
\thanks{S. K. Mohammed is also associated with the Bharti School of Telecom. Tech. and Management (BSTTM), IIT Delhi. The work of S. K. Mohammed was supported in part by the Jai Gupta Chair at I.I.T. Delhi.}
}



\maketitle

\begin{abstract}
Linear time-varying (LTV) systems model radar scenes where each reflector/target applies a delay, Doppler shift and complex amplitude scaling to a transmitted waveform. The receiver processes the received signal using the transmitted signal as a reference. The self-ambiguity function of the transmitted signal captures the cross-correlation of delay and Doppler shifts of the transmitted waveform. It acts as a blur that limits resolution, at the receiver, of the delay and Doppler shifts of targets in close proximity. This paper considers resolution of multiple targets and compares performance of traditional chirp waveforms with the Zak-OTFS waveform. The self-ambiguity function of a chirp is a line in the delay-Doppler domain, whereas the self-ambiguity function of the Zak-OTFS waveform is a lattice. The advantage of lattices over lines is better localization, and we show lattices provide superior noise-free estimation of the range and velocity of multiple targets. When the delay spread of the radar scene is less than the delay period of the Zak-OTFS modulation, and the Doppler spread is less than the Doppler period, we describe how to localize targets by calculating cross-ambiguities in the delay-Doppler domain. We show that the signal processing complexity of our approach is superior to the traditional approach of computing cross-ambiguities in the continuous time / frequency domain. 
\end{abstract}


\section{Introduction}
In 1953, Philip Woodward \cite{radar3} suggested that we view a radar scene as an unknown operator parameterized by delay and Doppler, and that we view radar waveforms as questions that we ask the operator. Woodward proposed to define a good question in terms of lack of ambiguity in the answer, he looked for waveforms with good resolution in delay and Doppler, and he suggested using a train of narrow TD pulses Gaussian pulses modulated with a broad Gaussian envelope. Woodward was not aware of Zak-OTFS modulation in 1953, but the waveform he proposed is strikingly similar to the TD realization of a Zak-OTFS carrier waveform, namely a pulse in the delay-Doppler (DD) domain. In this paper we consider resolution of multiple targets, and we compare performance of traditional chirp waveforms with the Zak-OTFS waveform.

We follow Woodward in connecting the problem of estimating target location in radar with the more challenging problem of precisely identifying the linear time-varying (LTV) system that describes the radar scene. It is well known that it is possible to identify an LTV system if and only if the area $A$ of the support of the channel DD spreading function satisfies $A < 1$ (see \cite{P}, \cite{PW} and \cite{HB}). Identification is possible if and only if the physical channel is \emph{underspread}. 

Section \ref{secsysmodel} describes how LTV systems model multipath channels where each reflector applies a delay, Doppler shift and complex amplitude scaling to the transmitted waveform. We identify the reflector by the path length  (or, through the speed of light, by the path delay), and by the Doppler shift determined by the radial velocity. We assume that the bandwidth of the waveform is small with respect to the carrier frequency so that a Doppler shift is accurately modeled by a frequency shift \cite{Bello63}. We process the received signal $y(t)$ using the transmitted waveform $x(t)$ as a reference. 

The self-ambiguity function $A_{x,x}(\tau, \nu)$ captures the cross-correlation of delay and Doppler shifts of the transmitted waveform. We view it as a blur that limits resolution of the delay and Doppler shifts of targets in close proximity. Moyal's Identity \cite{WMoran2001} places a lower bound on the volume under the squared ambiguity surface in terms of the energy of the transmitted waveform. Radar engineers shape self-ambiguity functions by designing waveforms that move the unavoidable volume under the squared ambiguity surface to regions where it matters least to the operational task of the radar.

Section \ref{chirpsec} begins by recalling that the self-ambiguity function of a chirp is supported on a line. A Gaussian pulse shaping filter transforms this line into a band with delay spread roughly $1/B$, and Doppler spread roughly $1/T$, where $B$ is the bandwidth and $T$ is the time of transmission.  Section \ref{chirpsec} then describes how to locate a single target located at $(\tau_1, \nu_1)$ by transmitting an up-chirp with slope $\alpha$ in the first $T/2$ seconds and a down-chirp with slope $\beta$ in the subsequent $T/2$ seconds. The method uses the cross-ambiguity function $A_{y,x}(\tau, \nu)$  between the received signal $y(t)$ and the transmitted signal $x(t)$, and the radar receiver computes $A_{y,x}(\tau, \nu)$ separately for each interval. In the first $T/2$ seconds the cross-ambiguity is supported on a band around the line $\nu - \alpha \tau = \nu_1 - \alpha \tau_1$, and in the subsequent $T/2$ seconds it is supported on a band around the line $\nu - \beta \tau = \nu_1 - \beta \tau_1$. When $\alpha$ and $\beta$ are distinct, the intersection of the lines/bands provides an estimate of the target location $(\tau_1, \nu_1)$. Section \ref{chirpsec} then considers estimation of four targets using the same pair of chirps. The cross-ambiguity function now exhibits four bands with positive slope and four bands with negative slope. There are intersection points that do not correspond to targets, and we refer to these intersection points as ghosts. Section \ref{chirpsec} concludes by describing a special case ($K=2$) of a general method for separating true targets from ghost targets by transmitting $K$ pairs of up-chirp/down-chirp waveforms \cite{Bajwa}. The disadvantage of this method is a loss in Doppler resolution because each individual chirp is now transmitted for only $T/2K$ seconds. 

Section \ref{sec4} introduces the Zak-OTFS carrier waveform as a pulse in the DD domain, a quasi-periodic localized function defined by a delay period $\tau_p$ and a Doppler period $\nu_p = 1/ \tau_p$. In previous work on wireless communications \cite{zakotfs1, zakotfs2}, we have shown that the Zak-OTFS Input/Output (I/O) relation is predictable and non-fading when the delay spread of the physical channel is less than the delay period $\tau_p$ and the Doppler spread is less than the Doppler period $\nu_p$. We refer to this condition as the crystallization condition and we have argued that a communication system should operate within this crystallization regime. The crystallization condition is more restrictive than the underspread condition, so it is always possible to identify the corresponding LTV operator. Section \ref{sec4} derives the self-ambiguity function for the Zak-OTFS waveform. The high intensity regions are localized at the points of the period lattice $\Lambda_p$ given by
\begin{eqnarray*}
    \Lambda_p & \Define & \{ (n \tau_p, m \nu_p) \, \vert \, n,m \in {\mathbb Z} \}.
\end{eqnarray*}

Since we transmit a single waveform for $T$ seconds, the high intensity regions have spread $1/B$ in delay and $1/T$ in Doppler. We do not lose Doppler resolution by subdividing $T$. Again, we use the peaks of the cross-ambiguity function $A_{y,x}(\tau, \nu)$ to estimate target locations. 

The TD realization of a Zak-OTFS waveform is a train of narrow pulses, exhibiting sharp peaks at the pulse locations. The peak to average power ratio (PAPR) is about $15$ dB which is not attractive for radar applications since transmission may require the use of highly linear power amplifiers which are typically power inefficient. However, it is possible to construct a spread waveform for which the PAPR of the TD realization is about $6$ dB by applying a type of discrete spreading filter to a pulse in the DD domain (for details, see \cite{zakotfs3}). The effect of the spreading filter is to distribute energy equally across all degrees of freedom in the DD domain. The effect of spreading in the TD is to produce a noise-like waveform that is well suited to radar applications. The ambiguity function of the spread waveform is supported on a lattice $\Lambda$ obtained by rotating the period lattice $\Lambda_p$. In this paper we aim to demonstrate the advantages of using a waveform with a self-ambiguity function that is supported on a lattice rather than a line. For clarity of exposition, we therefore focus on the original Zak-OTFS waveform rather than the spread waveform.

In Section \ref{simsec} we first consider noise-free estimation of the range and velocity of multiple targets using a Zak-OTFS waveform and two different chirp waveforms. We consider four targets uniformly distributed in six rectangles $\Omega_i = [0, (7-i)] \mu s \, \times \,  [-200(7-i) \, , \, 200(7-i)]$ Hz,
$i=1,2,\cdots, 6$. The targets are clustered and target spacing decreases as $i$ increases. Estimated range is half the product of the target delay and the speed of light. Estimated velocity is half the product of the target Doppler shift and the RF carrier wavelength (carrier frequency is $1$ GHz). The numerical simulations illustrate that the Zak-OTFS waveform provides superior range and velocity estimation compared to chirps. We conclude Section \ref{simsec} by considering range and velocity estimation accuracy as a function of SNR. At low SNR, the estimation error is almost identical for chirp and Zak-OTFS waveforms, but with increasing SNR, the error performance of Zak-OTFS is significantly better.  

\begin{table}
\centering
\caption{Comparison with prior works on identification of linear time-varying systems. $B$ and $T$ are the waveform bandwidth and time-duration respectively. $K$ is the maximum possible number of resolvable targets. $\Delta \tau$ and $\Delta \nu$ are the delay and Doppler domain resolution respectively. The delay and Doppler shift corresponding to the $i$-th target are denoted by $\tau_i$ and $\nu_i$ respectively.}
\begin{tabular}{|c||c|c|c|}
\hline
    Identification  & Max. resolvable & Complexity & Resolution \\
    approach &  targets ($K$) & & \\
    \hline 
    Herman-Strohmer  & $\sqrt{BT}$ & $O(K^3)$ & $\Delta \tau \propto \frac{1}{B}$ \\
    \cite{HS} & & & $\Delta \nu \propto \frac{1}{T}$ \\
\hline
    Friedlander \cite{F} & $\sqrt{BT}$ & $O(K^3)$ & Infinite \\
    Bajwa et. al \cite{BGE} & & &  \\
    \hline
    Harms et. al.  & $\sqrt{BT}$ & $O(K^3)$ & Infinite \\
    \cite{HBC1, HBC2, Bajwa} & & & \\
    \hline
    DD domain & & & \\
    Cross-ambiguity & $\frac{BT}{4}$ & $(BT)^2$ & $\Delta \nu \propto \frac{4}{T}$ \\
    Chirp (LFM) pulses & & & \\
    Section \ref{chirpsec} (this paper) & & & $\Delta \tau \propto \frac{1}{B}$ \\
    \hline
    DD domain  & & & \\
    Cross-ambiguity & $BT$ & $BT \, \log(BT)$ & $\Delta \nu \propto \frac{1}{T}$ \\
    Zak-OTFS waveform & & & \\
    Section \ref{sec4} (this paper) & & & $\Delta \tau \propto \frac{1}{B}$ \\   
    \hline
\end{tabular}
\label{tabcmp}
\end{table}

{\textbf{Comparison with Prior Work:} We compare our proposed identification method with prior works in the summary presented in Table~\ref{tabcmp}. The first entry refers to a method based on compressed sensing that is able to identify a target located on discrete grid in the DD domain \cite{HS}. The disadvantage of this method is that targets need not live on a grid, and mismatch to the discretized basis may degrade performance \cite{CSPC}. The second entry refers to parametric techniques that have been proposed as an alternative to matched-filter processing \cite{F}, \cite{BGE}. These approaches employ sequential recovery of the delay shift followed by the Doppler shift, or vice-versa, and the drawback is error propagation between stages. The third entry refers to the method of transmitting multiple pairs of up- and down-chirps to resolve multiple targets \cite{HBC1}, \cite{HBC2}, \cite{Bajwa}. The fourth and fifth entries refer to a method where
the target locations are given by the peak of the cross-ambiguity between the transmitted and the received waveform. Although the parametric methods have better delay Doppler resolution when compared to the cross-ambiguity based method, the number of
resolvable targets $K$ is significantly smaller when compared to that for the cross-ambiguity based method.
A novel aspect of our paper is that, instead of traditional time/frequency domain based computation of the cross-ambiguity 
function, we propose to compute the cross-ambiguity in the DD domain. Traditional time/frequency domain computation of cross-ambiguity has complexity $O((BT)^2)$. However, the complexity can be smaller if computed in the DD domain. For the Zak-OTFS waveform the complexity is only $BT \, \log (BT)$ (see discussion in Section \ref{sec4}). This is because, the self-ambiguity of the Zak-OTFS waveform is not spread but is instead localized to a lattice. For the chirp waveform the complexity is higher since its self-ambiguity function is not localized but is spread along a line in the DD domain. The chirp waveform (fourth entry in the table) suffers from the problem of ghost targets as discussed above and therefore multiple chirp pulses (at least four pulses) need to be transmitted to resolve the true targets, which degrades the Doppler domain resolution from $1/T$ to $4/T$, and due to which the maximum number of resolvable targets is four times smaller when compared to that for the Zak-OTFS waveform. The issue of ghost targets does not arise when we use the Zak-OTFS waveform since its self-ambiguity function is supported on a lattice, and hence its Doppler domain resolution does not suffer.

\section{System Model}
\label{secsysmodel}
An LTV system is an operator that acts on a probe waveform. The response is a sum of $P$ copies of the probe waveform, where the $i$th copy is delayed by $\tau_i$, shifted in frequency by $\nu_i$ and scaled by a complex number $h_i$. each triple $(\tau_i, \nu_i, h_i)$ is associated with a target, and our objective is to identify the LTV operator by identifying every triple.
The received signal is given by 
\begin{eqnarray}
\label{otfsrxsigneqn}
y(t)=\int \int h(\tau, \nu) \, x(t-\tau) \, e^{j2\pi \nu (t-\tau)} d\tau \, d\nu +n(t)
\end{eqnarray}
where $h(\tau, \nu)$ is the delay-Doppler (DD) spreading function specifying the physical channel between the transmitter and the receiver and $n(t)$ is additive white Gaussian noise (AWGN) at the receiver. We write
\begin{eqnarray}
\label{heqn}
h(\tau,\nu) &=& \sum_{i=1}^{P} h_i \delta(\tau-\tau_i) \delta(\nu-\nu_i),
     \end{eqnarray}
where $\delta(.)$ denotes the Dirac-delta function. Substituting (\ref{heqn}) in (\ref{otfsrxsigneqn}), we obtain
     \begin{eqnarray}
     \label{rxsignaleqn}
     y(t) &=& \sum_{i=1}^{P} h_i e^{j 2\pi \nu_i(t-\tau_i)}x(t-\tau_i) + n(t).
     \end{eqnarray}
When there is a single target ($P=1$), the maximum likelihood estimate (MLE) for the delay shift $\tau_1$ and Doppler shift $\nu_1$ is given by matched filtering (see Apppendix \ref{app1} for details).
\begin{eqnarray}
 \label{MLEeqn}
 (\hat{\tau_1},\hat{\nu_1}) &\overset{\Delta}{=}& \underset{\tau,\nu}{argmax} \left | A_{y,x}(\tau, \nu)\right|.
      \end{eqnarray}
Here $A_{y,x}(\tau, \nu)$ denotes the cross-ambiguity between between the probe waveform $x(t)$ and the received waveform $y(t)$, which is given by
\begin{eqnarray} \label{cross_ambiguity_eqn}
     A_{y,x}(\tau, \nu) = \int y(t) x^*(t-\tau) e^{-j2\pi\nu (t-\tau)}dt. 
     \end{eqnarray}
We now rewrite (\ref{cross_ambiguity_eqn}) in terms of
\emph{twisted convolution} in the DD domain. The role of twisted convolution in linear time-varying systems is identical to that of linear convolution in linear time-invariant systems. The twisted convolution of DD domain functions $a(\tau, \nu)$ and $b(\tau, \nu)$ is given by 
\begin{eqnarray}
\label{twistconv}
    a(\tau, \nu) \ast_{\sigma} b(\tau, \nu) & & \nonumber \\
    & & \hspace{-28mm} = \iint a(\tau', \nu') b(\tau - \tau', \nu - \nu') \, e^{j 2 \pi \nu' (\tau - \tau')} \, d\tau' \, d\nu'.
\end{eqnarray} Note that twisted convolution is associative (just like linear convolution) but not commutative (for
more information about twisted convolution, see \cite{zakotfs1, zakotfs2, zakotfs3, zakotfsbook}). We rewrite (\ref{cross_ambiguity_eqn}) in the form
\begin{eqnarray}
    \label{sim_cross_ambiguity_eqn}
 A_{y,x}(\tau, \nu) =  h(\tau, \nu) \ast_{\sigma} A_{x,x}(\tau, \nu)\nonumber \\
 &\hspace{-3.5cm} +\int n(t) x^*(t-\tau) e^{-j2\pi\nu (t-\tau)}dt,
  \end{eqnarray} and refer the reader to Appendix \ref{app2})
  for more details. The function $A_{x,x}(\tau, \nu)$
  appearing in (\ref{sim_cross_ambiguity_eqn}) is the self- or auto-ambiguity function of the probe waveform $x(t)$.
  This function captures the cross correlation of delay and Doppler shifts of the probe signal, which we view as a blur
  that limits resolution of the delay and Doppler shifts of targets in close proximity \cite{radar3}. It is given by
\begin{eqnarray}
\label{self_ambiguity_eqn}
A_{x,x}(\tau, \nu)=\int x(t) x^*(t-\tau) e^{-j2\pi\nu (t-\tau)}dt.
\end{eqnarray}
Moyal's Identity \cite{WMoran2001} places a lower bound on the volume under the squared ambiguity surface as a function of the energy $E_T$ of the probe waveform.
\begin{eqnarray}
    \iint \left\vert A_{x,x}(\tau, \nu) \right\vert^2 \, d\tau \, d\nu & = & E_T^2.
\end{eqnarray}It encapsulates, in a slightly different form,
the Hiesenberg Uncertainty Principle, and places fundamental
limits on target resolution. We have
\begin{eqnarray}
\label{paper_eqn23}
    \left\vert A_{x,x}(\tau, \nu) \right\vert & = & \left\vert  \int x(t) x^*(t-\tau) e^{-j2\pi\nu (t-\tau)}dt \right\vert  \nonumber \\
    &  &  \hspace{-20mm} \leq  \,  \int \vert x(t) x^*(t-\tau) e^{-j2\pi\nu (t-\tau)} \vert \, dt \, = \,  E_T.
\end{eqnarray}where
\begin{eqnarray}
    A_{x,x}(0,0) & = & \int \vert x(t) \vert^2 \, dt  \, = \, E_T.
\end{eqnarray} The peak value of the self-ambiguity function $A_{x,x}(\tau, \nu)$ occurs at $(\tau, \nu) = (0,0)$.
When there is a single target, we have
\begin{eqnarray}
\label{ayxp1}
    A_{y,x}(\tau, \nu) & = & h_1 \, {\Big (} \delta(\tau - \tau_1) \, \delta(\nu - \nu_1) {\Big )} \, \ast_{\sigma} \, A_{x,x}(\tau, \nu) \nonumber \\
    & & \, + \, \int n(t) x^*(t-\tau) e^{-j2\pi\nu (t-\tau)}dt.
\end{eqnarray}The signal term in (\ref{ayxp1}) is
simply the self-ambiguity function $A_{x,x}(\tau, \nu)$ shifted by $(\tau_1, \nu_1)$ in the DD domain.
\begin{eqnarray}
\label{singlepathshift}
    h_1 \, {\Big (} \delta(\tau - \tau_1) \, \delta(\nu - \nu_1) {\Big )} \, \ast_{\sigma} \, A_{x,x}(\tau, \nu) & & \nonumber \\
    & & \hspace{-50mm} = h_1 A_{x,x}(\tau-\tau_1, \nu-\nu_1)e^{j 2 \pi \nu_{1}(\tau-\tau_1)},
\end{eqnarray}The peak value of $\left\vert A_{x,x}(\tau- \tau_1, \nu - \nu_1) \right\vert$ occurs at $(\tau_1, \nu_1)$, and in the absence of noise, the peak of the absolute value of the cross-ambiguity function $A_{y,x}(\tau, \nu)$ also occurs at $(\tau_1, \nu_1)$. In the presence of noise, this peak may shift, and the shape/spread of the self-ambiguity function $A_{x,x}(\tau, \nu)$ determines the accuracy of estimation.

The shape/spread of the self-ambiguity function $A_{y,x}(\tau, \nu)$ limits the ability to separate multiple targets. For example, the self-ambiguity function of the linear frequency modulated (LFM) pulse introduced in Section \ref{chirpsec} is characterized by a line of large intensity that couples delay and Doppler shifts. Two targets described by parameters that fall on this line are indistinguishable. However, it is still possible to separate multiple targets by probing the LTV operator with a diverse set of LFM waveforms
\cite{Bajwa}, and this is described in Section \ref{chirpsec}.

The Zak-OTFS carrier waveform is a pulse in the DD domain, and in Section \ref{sec4} we observe that the self-ambiguity function is characterized by regions of high intensity that form a discrete lattice rather than a line (see also \cite{zakotfs2, zakotfsbook}). We then demonstrate that there are advantages to choosing a probe waveform with these characteristics.

When choosing a probe waveform for a radar application,
the regions of high intensity in the self-ambiguity function should correspond to target delay-Doppler locations of low operational significance.



\subsection{Delay-Doppler signal processing}
We begin by describing the Zak-transform (${\mathcal Z}_t$) which expresses the DD representation
$x_{\mbox{\scriptsize{dd}}}(\tau, \nu)$ of a
signal $x$ in terms of the TD representation $x(t)$.
The Zak transform is parameterized by a delay period
$\tau_p$ for which the corresponding Doppler period
$\nu_p = 1/\tau_p$. It is given by
\begin{eqnarray}
\label{zaktransformeqn}
    x_{\mbox{\scriptsize{dd}}}(\tau, \nu) & \hspace{-2.5mm} = & \hspace{-2.5mm} {\mathcal Z}_t{\Big (} x(t)  {\Big )} = \sqrt{\tau_p} \sum\limits_{k \in {\mathbb Z}} \hspace{-1mm} x(\tau + k \tau_p) \, e^{-j 2 \pi \nu k \tau_p}.
\end{eqnarray}
See \cite{Zak67, Janssen88} for more details and see Chapter $2$ of \cite{zakotfsbook} for a comprehensive introduction. Observe that
\begin{eqnarray}
     x_{\mbox{\scriptsize{dd}}}(\tau + n\tau_p, \nu + m \nu_p) & = & e^{j 2 \pi n \nu \tau_p} \, x_{\mbox{\scriptsize{dd}}}(\tau , \nu ),
\end{eqnarray}for all $n,m \in {\mathbb Z}$. The DD realization $x_{\mbox{\scriptsize{dd}}}(\tau, \nu)$
is quasi-periodic, with period $\tau_p$ along the delay axis, and period $\nu_p$ along the Doppler axis. Conversely, the TD representation $x(t)$ of a quasi-periodic function $x_{\mbox{\scriptsize{dd}}}(\tau, \nu)$ in the DD domain is given by the inverse Zak transform
\begin{eqnarray}
\label{invzakt}
    x(t) & = & {\mathcal Z}_t^{-1}{\Big (} x_{\mbox{\scriptsize{dd}}}(\tau , \nu ) {\Big )} \, = \, \sqrt{\tau_p} \int\limits_{0}^{\nu_p} x_{\mbox{\scriptsize{dd}}}(t , \nu ) \, d\nu.
\end{eqnarray} Note that there is also a frequency-domain (FD) transform ${\mathcal Z}_f$ that expresses the DD representation $x_{\mbox{\scriptsize{dd}}}(\tau , \nu )$ in terms of the FD representation of the signal $x$ (see Chapter $2$ of \cite{zakotfsbook}).

Appendix \ref{app3} shows that the Zak transform is a unitary transformation that preserves
inner products. Given quasi-periodic functions $a_{_{\mbox{\footnotesize{dd}}}}(\tau , \nu), b_{_{\mbox{\footnotesize{dd}}}}(\tau , \nu)$ and their corresponding TD realizations $a(t), b(t)$ we have
\begin{eqnarray}
\label{ch2_eqninnerproductdd}
  \int\limits_{-\infty}^{\infty} \, a(t) \, b^*(t) \, dt & \hspace{-2.5mm} = & \hspace{-2.5mm} \int\limits_{0}^{\tau_p} \int\limits_{0}^{\nu_p} a_{_{\mbox{\footnotesize{dd}}}}(\tau , \nu) \, b^*_{_{\mbox{\footnotesize{dd}}}}(\tau , \nu) \, d\nu \, d\tau.
\end{eqnarray}
We now describe how cross-ambiguity is the generalization to the DD domain of cross-correlation in TD. Appendix \ref{app4} shows that the DD realization of the TD signal $y(t)$ is related to the DD realization of the probe signal $x(t)$ by
\begin{eqnarray}
    \label{iorelation}
    y_{\mbox{\scriptsize{dd}}}(\tau, \nu) & = & h(\tau, \nu) \, \ast_{\sigma} \,  x_{\mbox{\scriptsize{dd}}}(\tau, \nu) \, + \, n_{\mbox{\scriptsize{dd}}}(\tau, \nu)
\end{eqnarray}where $n_{\mbox{\scriptsize{dd}}}(\tau, \nu)$ denotes AWGN in the DD domain. Appendix \ref{app5} shows that the cross-ambiguity $A_{y,x}(\tau, \nu)$
can be evaluated in the DD domain, and is given by
\begin{eqnarray}
\label{crossambigdd}
    A_{y,x}(\tau, \nu) &  &  \nonumber \\
    & & \hspace{-25mm} = \int\limits_{0}^{\tau_p} \hspace{-1mm} \int\limits_{0}^{\nu_p} \hspace{-1mm} y_{\mbox{\scriptsize{dd}}}(\tau', \nu') \, x_{\mbox{\scriptsize{dd}}}^*(\tau' - \tau, \nu' - \nu) \, e^{-j 2 \pi \nu (\tau' - \tau)} \, d\tau' \, d\nu'.
\end{eqnarray}It is evident from the integrand that the cross-ambiguity $A_{y,x}(\tau, \nu)$ is simply the
correlation between the received DD signal $y_{\mbox{\scriptsize{dd}}}(\tau', \nu')$ and the probe signal $x_{\mbox{\scriptsize{dd}}}(\tau', \nu')$ shifted by $(\tau, \nu)$, which is simply
\begin{eqnarray}
    \delta(\tau' - \tau)\delta(\nu' - \nu) \ast_{\sigma} x_{\mbox{\scriptsize{dd}}}(\tau', \nu') & &  \nonumber \\
    & & \hspace{-30mm} = x_{\mbox{\scriptsize{dd}}}(\tau' - \tau, \nu' - \nu) \, e^{j 2 \pi \nu (\tau' - \tau)}. 
\end{eqnarray}

\section{Chirp signals}
\label{chirpsec}
A chirp signal takes the form
\begin{eqnarray}
    \label{chirp signal}
c(t) = e^{j \pi a t^2}, -\infty < t < \infty,
\end{eqnarray}
where $a$ is real and is referred to as the \emph{slope} of the chirp signal $c(t)$. Observe that $c(t)$ is not limited in time and bandwidth.

\subsection{Pulse shaping to limit time and bandwidth}
The Zak transform relates the TD representation $c(t)$ to the DD domain representation $c_{\mbox{\scriptsize{dd}}}(\tau, \nu)$.
\begin{eqnarray}
    \label{zak chirp}
    c_{\mbox{\scriptsize{dd}}}(\tau, \nu) = {\mathcal Z}_t(c(t)).
\end{eqnarray}The pulse shaping filter
\begin{eqnarray}
   \label{filter eqn} 
w(\tau, \nu) & = & w_1(\tau) \, w_2(\nu), \nonumber \\
w_1(\tau) & \Define &  \left(\frac{2 \alpha B^2}{\pi}\right)^{1/4} e^{-\alpha B^2 \tau^2} \nonumber \\
w_2(\nu) & \Define &  \left(\frac{2 \beta T^2}{\pi}\right)^{1/4} e^{-\beta T^2 \nu^2}
\end{eqnarray}limits the time duration to $T$ seconds
and limits the bandwidth to $B$ Hz (see \cite{zakotfs1, zakotfs2, zakotfsbook, Gausspaper} for details).
The filtered chirp signal $u_{\mbox{\scriptsize{dd}}}(\tau, \nu)$ is simply the twisted convolution of the pulse-shaping filter with the chirp signal.
\begin{eqnarray}
    \label{chirp filter}
     u_{\mbox{\scriptsize{dd}}}(\tau, \nu) = w(\tau, \nu) \ast_{\sigma} c_{\mbox{\scriptsize{dd}}}(\tau, \nu).
\end{eqnarray}
The TD realization $u(t)$ is time and bandwidth limited to $T$ seconds, bandwidth-limited to $B$ Hz, and is given by
\begin{eqnarray}
    \label{TD Chirp limted}
     u(t) = {\mathcal Z_{t}}^{-1} {\Big (} u_{\mbox{\scriptsize{dd}}}(\tau, \nu) {\Big )}.
\end{eqnarray}From \cite{Gausspaper} it follows that
\begin{eqnarray}
     u(t) & = & {\mathcal Z_{t}}^{-1} {\Big (} (w_1(\tau) w_2(\nu)) *_{\sigma} {\mathcal Z}_t(c(t)) {\Big )} \nonumber \\
     & = & w_1(t) \star \left[ W_2(t) \, c(t) \right], \nonumber \\
     W_2(t) & \Define & \int w_2(\nu) \, e^{j 2 \pi \nu t} \, dt,
\end{eqnarray}where $\star$ denotes linear convolution.

\subsection{Ambiguity functions}
The self-ambiguity function of our initial chirp $c(t)$ is supported on the line $\nu = a \tau$ in the DD domain,
\begin{eqnarray} \label{ct_ambig}
     A_{c,c}(\tau, \nu) = \int c(t) \,\,c^*(t-\tau) e^{-j2\pi\nu (t-\tau)}dt \nonumber \\
     &\hspace{-5cm}= e^{-j \pi a \tau^2}\, e^{j 2 \pi \nu \tau} \int e^{-j 2 \pi (\nu -a\tau)t} dt \nonumber \\
     &\hspace{-6cm}=e^{j 2 \pi (\nu \tau-\frac{a \tau^{2}}{2})} \,\,\,\, \delta(\nu-a\tau) \nonumber \\
     &\hspace{-6cm}= e^{j \pi \nu \tau} \, \delta(\nu - a \tau).
     \end{eqnarray}
     
\begin{lemma}
\label{lm1}
    The self-ambiguity function $A_{u,u}(\tau, \nu)$ of the filtered chirp $u(t)$ is given by
 
\begin{eqnarray}
\label{ambiguityuueqn}
A_{u,u}(\tau, \nu) = w(\tau, \nu) \ast_{\sigma} A_{c,c}(\tau, \nu) \ast_{\sigma} w_{\mbox{\tiny{mf}}}(\tau, \nu),
\end{eqnarray}
where $w(\tau, \nu)$ is given by (\ref{filter eqn}), $A_{c,c}(\tau, \nu)$ is the auto-ambiguity function of $c(t)$, and $w_{\mbox{\tiny{mf}}}(\tau, \nu) = w^{*}(-\tau, -\nu) \,\,e^{j 2 \pi \tau \nu}$ is the matched filter.
\end{lemma}
\begin{IEEEproof}
See Appendix \ref{appendix_lm1}.
\end{IEEEproof}This is a special case of a more general result. The proof given in Appendix \ref{appendix_lm1}
shows that when we apply an arbitrary pulse shaping filter $w(\tau, \nu)$ to an arbitrary quasi-periodic
function $c_{\mbox{\scriptsize{dd}}}(\tau, \nu)$ to obtain a quasi-periodic function $u_{\mbox{\scriptsize{dd}}}(\tau, \nu)$, then the self-ambiguity function is given by (\ref{ambiguityuueqn}).

\begin{theorem}
    \label{thm1}
    The self-ambiguity function $A_{u,u}(\tau, \nu)$ of the filtered chirp $u(t)$ is given by (\ref{Auufinaleqn}) (see top of this page).
\end{theorem}  
\begin{IEEEproof}
See Appendix \ref{appendix_thm1}.
\end{IEEEproof} 


\begin{figure*}
{\vspace{-7mm}
\small
\begin{eqnarray}
\label{Auufinaleqn}
A_{u,u}(\tau, \nu) & \hspace{-2mm} =&  \hspace{-2mm} \left(\frac{2 \alpha B^2}{\pi}\right)^{1/2} 
 \, \left(\frac{2 \beta T^2}{\pi}\right)^{1/2} \, \sqrt{\frac{\pi}{2\beta T^2}} \,\sqrt{\frac{\pi}{2\alpha B^2}} \, e^{-\frac{\pi^2 \nu^2}{(2\alpha B^2)^2}} \, e^{j \pi \nu \tau}  \int    e^{-\frac{(a\Tilde{\tau}-\nu)^2}{2}}\,e^{-\frac{\pi^2 \Tilde{\tau}^2}{(2\beta T^2)^2}} e^{-j 2\pi a \Tilde{\tau}^2} \, e^{-\frac{\alpha B^2(\tau-\Tilde{\tau})^2}{2}} \, d\Tilde{\tau}.
\end{eqnarray}
\normalsize}
\hrulefill
\end{figure*}

Observe that the main lobe of the Gaussian pulse shaping filter $w(\tau, \nu) $
given by (\ref{filter eqn}) has a spread of roughly $1/B$ along the delay axis, and a spread of roughly $1/T$ Hz along the Doppler axis. The same is true for the matched filter $w_{\mbox{\tiny{mf}}}(\tau, \nu)$. 
The self-ambiguity function $A_{c,c}(\tau, \nu)$
of the initial chirp is supported on a line, and pulse shaping transforms this line into a band with delay spread roughly $1/B$ and Doppler spread roughly $1/T$. This is illustrated in Fig.~\ref{chirp working}
which displays the intensity of the cross-ambiguity
function for two filtered chirp signals.


\subsection{Locating a single target}
We locate a single target located at $(\tau_1, \nu_1)$
by transmitting an up-chirp with slope $\alpha$ in the first $T/2$ seconds and a down-chirp with slope $\beta$ in the subsequent $T/2$ seconds. The radar receiver computes the cross-ambiguity $A_{y,x}(\tau, \nu)$ separately for each interval as described in (\ref{crossambigdd}). In the first $T/2$ seconds
the cross-ambiguity is supported on a band around the line $\nu - \alpha \tau = \nu_1 - \alpha \tau_1$
and in the subsequent $T/2$ seconds it is supported on a band around the line $\nu - \beta \tau = \nu_1 - \beta \tau_1$. 

For single target scenarios, we divide the total time duration into two halves, $T/2$ seconds each. In the first half we transmit a chirp $u(t)$ (of duration $T/2$ and bandwidth $B$) with positive slope $\alpha_1$ and for the remaining half we transmit a chirp (of duration $T/2$ and bandwidth $B$) with negative slope $\beta_1$. The radar receiver computes the cross-ambiguity $A_{y,u}(\tau, \nu)$ separately for each half.
From (\ref{singlepathshift}) we know that the magnitude of cross-ambiguity (noise-free component) is $\vert A_{u,u}(\tau - \tau_1, \nu - \nu_1) \vert$, i.e., $A_{u,u}(\tau, \nu)$ translated/shifted by $(\tau_1, \nu_1)$. Since the support of $A_{u,u}(\tau, \nu)$ is on/around the line $\nu - a \tau = 0$ (when slope is $a$), the support for $\vert A_{u,u}(\tau - \tau_1, \nu - \nu_1) \vert$ is on/around the line $\nu - a \tau = \nu_1 - a \tau_1$. Fig.~\ref{chirp working} illustrates that when $\alpha$ and $\beta$ are distinct, the lines intersect only at the target location $(\tau_1, \nu_1).$   

\begin{figure}
\hspace{-5mm}
      \includegraphics[scale=0.14]{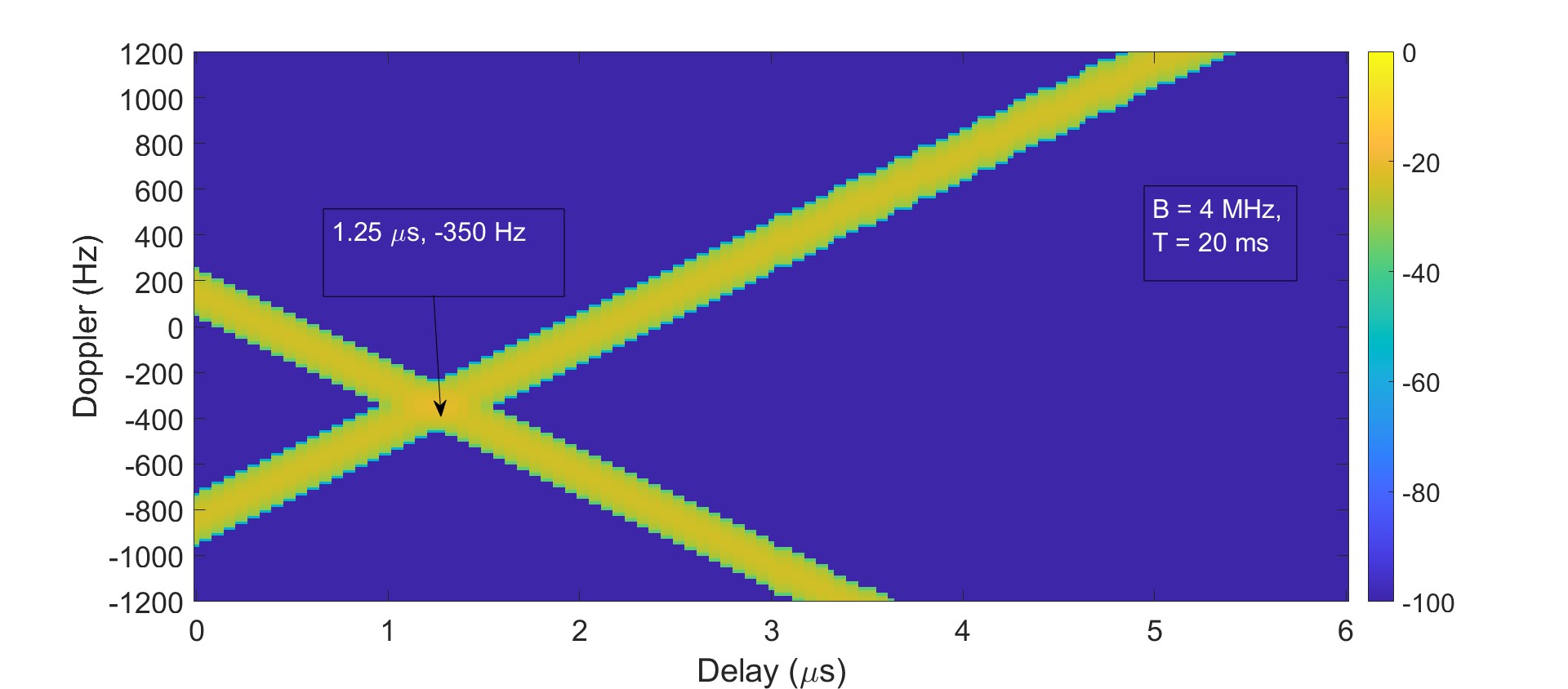}
  \caption{Heat map illustrating the intensity of the cross-ambiguity function for two filtered chirp signals, an up-chirp with slope
  $2B/T = 4 \times 10^8$ Hz$^2$ and a down-chirp with slope $-2B/T$. The bandwidth is $B = 4$ MHz, the time duration $T=20$ ms, and the target location is $\tau_1 = 1.25 \mu s$ and $\nu_1 = -350$ Hz.
  The up-chirp is transmitted in the first $T/2$ seconds, the down-chirp is
  transmitted in the subsequent $T/2$ seconds, and the radar receiver computes the cross-ambiguity separately for the two intervals.
  The Doppler spread of each band is roughly $2/T = 100$ Hz,
  and the delay spread is roughly $1/B = 0.25 \mu s$.
  The intensity pattern is typical of Gaussian pulse shaping.}
  \label{chirp working}
     \end{figure}

\subsection{Locating multiple targets}
We next consider four targets located at $(1 \mu s, -400 \mbox{\scriptsize{Hz}})$,
$(3.125 \mu s, 175 \mbox{\scriptsize{Hz}})$, $(2.375 \mu s, -550 \mbox{\scriptsize{Hz}})$
and $(4.25 \mu s, -600 \mbox{\scriptsize{Hz}})$
using the same pair of up- and down-chirps. Fig.~\ref{ghost in chirp} illustrates the heatmap of
the cross-ambiguity function for the two filtered chirp signals. We observe four bands/lines with positive slope and four bands with negative slope. The intersection points are marked by circles, where blue circles indicate true targets and red circles indicate non-targets (\emph{ghosts}).

It is possible to separate true targets from ghost targets by transmitting two distinct pairs of up-chirp/down-chirp signals \cite{Bajwa}. The method is to transmit an up-chirp with slope $\alpha_1$ in the first $T/4$ seconds and a down-chirp with slope $\beta_1$ in the next $T/4$ seconds. Then, in the remaining $T/2$ seconds, to transmit an up-chirp with slope $\alpha_2$ for $T/4$ seconds followed by a down-chirp with slope $\beta_2$ for $T/4$. Choosing slopes $\alpha_1 \ne \alpha_2$ and $\beta_1 \ne \beta_2$ results in ghost locations that are different in the two cross-ambiguity plots. While it is now possible to separate true targets from ghosts, there is a loss in resolution. This is because the Doppler spread of the bands in the cross-ambiguity plots has increased from $2/T$ to $4/T$, which might result in two or more targets being observed as a single target.

In the next Section, we avoid this trade-off between resolution and the ability to separate multiple targets by using a quasi-periodic pulse in the DD domain as a probe waveform. The self-ambiguity function is supported on a lattice in the DD domain, with lattice points separated by $\tau_p$ along the delay axis and by $\nu_p = 1/\tau_p$ along the Doppler axis. The lattice geometry will avoid the need to identify ghost targets, and the transmission of a single probe signal over $T$ seconds will avoid compromising Doppler resolution.

The heatmap of the cross-ambiguity function for both the received up-chirp and the down-chirp is plotted in
Fig.~\ref{ghost in chirp}.
\begin{figure}[tbh]
     \hspace{-7mm} \includegraphics[scale=0.14]{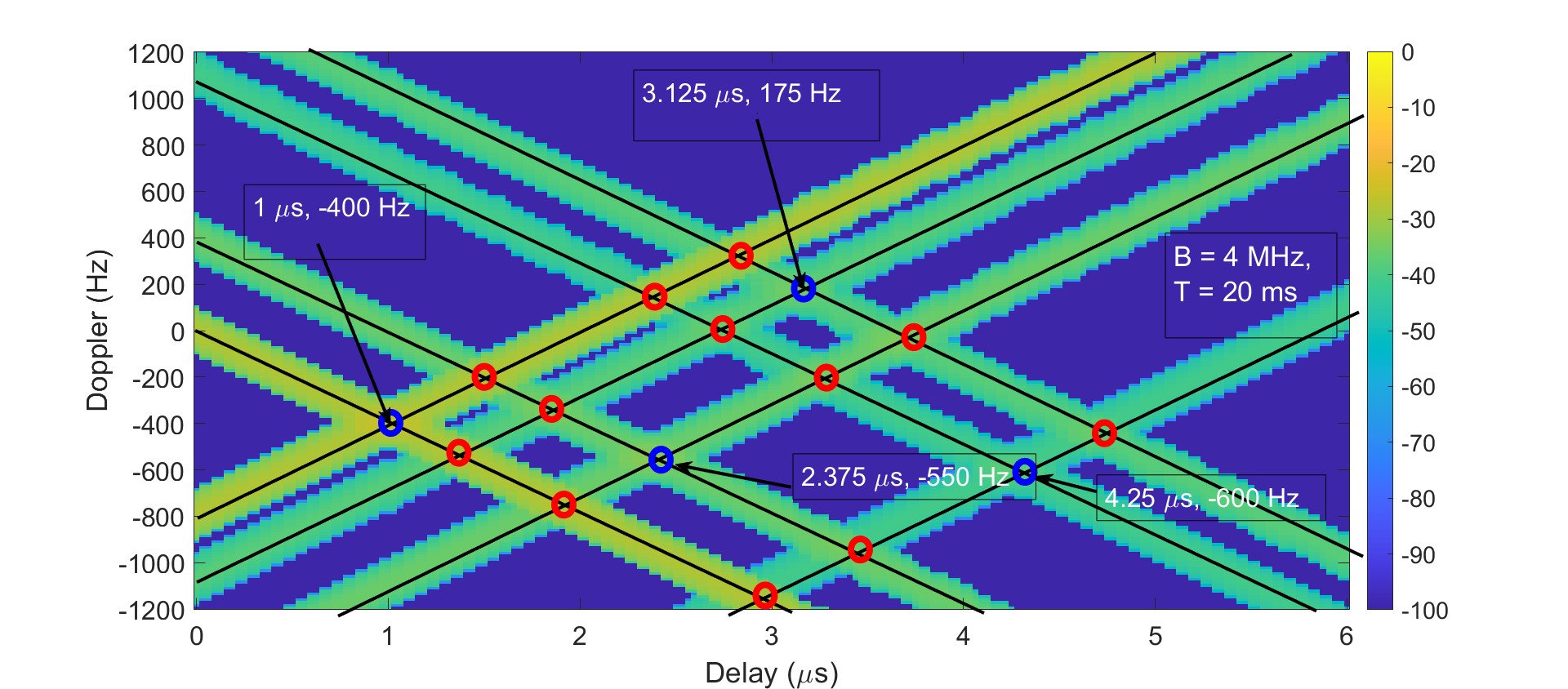}
  \caption{Heat map illustrating the intensity of the cross-ambiguity function for two filtered chirp signals, an up-chirp with slope $2B/T = 4 \times 10^8$ and a down-chirp with slope $2B/T$. The bandwidth is $4$ MHz, the time duration $T=20$ ms, and there are four targets located at $(1 \mu s, -400 \mbox{\scriptsize{Hz}})$,
$(3.125 \mu s, 175 \mbox{\scriptsize{Hz}})$, $(2.375 \mu s, -550 \mbox{\scriptsize{Hz}})$
and $(4.25 \mu s, -600 \mbox{\scriptsize{Hz}})$.
The up-chirp is transmitted in the first $T/2$ seconds,
the down-chirp is transmitted in the subsequent $T/2$ seconds, and the radar receiver computes the cross-ambiguity separately for the two intervals. There are four bands/lines with positive slope and four bands with negative slope. The intersection points are marked by
circles, where blue circles indicate true targets and red circles indicate non-targets (\emph{ghosts}).
The intensity pattern is characteristic of Gaussian pulse shaping.}
  \label{ghost in chirp}
     \end{figure}
Clearly, since there are four targets, there are four straight lines/bands with positive slope and another four with negative slope. However, these lines/bands intersect at more than four locations (marked with colored circles) out of which only four (blue circles) are the true targets and the others (red circles) are false (``ghost") targets.

With just an up-chirp and a down-chirp it is difficult to separate the ghost targets from the true targets. This issue was considered in \cite{Bajwa} and the solution proposed was to transmit two pairs of up-chirp/down-chirp signals, i.e.,
an up-chirp/down-chirp with slope $\alpha_1$ and $\beta_1 \ne \alpha_1$ respectively (each of duration $T/4$ seconds) followed by another up-chirp/down-chirp pair with different slope $\alpha_2 \ne \alpha_1$ and $\beta_2 \ne \alpha_2$ respectively (each of duration $T/4$ seconds).

Since the slopes for the two up-chirp/down-chirp pairs are different, the ghost locations in the cross-ambiguity plot of the first up-chirp/down-chirp pair are different from the ghost locations
in the plot for the second up-chirp/down-chirp pair, the only targets that are common to both are the true targets. Although this approach is able to separate true targets from ghost targets, it
suffers from loss of resolution along the Doppler domain since now the Doppler domain spread of the cross-ambiguity bands are $\approx 4/T$ which could result in two or more targets being observed as a single target.

In the next section, we therefore consider a new family of radar signals (based on the Zak-transform) which are essentially quasi-periodic pulses in the DD domain and for which the auto-ambiguity function is supported on a lattice in the DD domain with lattice points separated by $\tau_p$ along the delay
domain and separated by $\nu_p = 1/\tau_p$ along the Doppler domain. As we shall see later,
lattice-type auto-ambiguity does not suffer from the ghost target issue and has higher Doppler resolution as only one waveform of duration $T$ seconds is transmitted. 

\section{Zak-OTFS signals}
\label{sec4}
A pulse centered at $(\tau_0, \nu_0)$ in the DD domain
is a quasi-periodic localized function parameterized
by a delay period $\tau_p$ and a Doppler period $\nu_p = 1/\tau_p$ that is given by
   \begin{eqnarray} \label{DD localized signal}
     p_{_{\mbox{\scriptsize{dd}},\tau_0,\nu_0}}(\tau, \nu) & = & \sum\limits_{m \in \mathbb{Z}} \sum\limits_{n \in \mathbb{Z}}  {\Big [} e^{j2\pi \nu_{0} n\tau_{p}} \delta(\tau-n\tau_{p}-\tau_{0}) \nonumber \\
     & & \hspace{14mm} \delta(\nu-m\nu_{p}-\nu_{0}) {\Big ]}.
      \end{eqnarray}where $0 \leq \tau_0 < \tau_p$ and
      $0 \leq \nu_0 < \nu_p$ (see \cite{zakotfs1, zakotfsbook, Zak67, Janssen88}). We apply the inverse Zak
      transform to obtain the TD realization, which is
      an infinite train of pulses spaced $\tau_p$ apart, modulated by a tone with frequency $\nu_0$.
      Hence the name \emph{pulsone}. Note that the coordinate $\tau_0$ determines the offset of the pulse train.
      The constituent pulses are Dirac-delta functions, so the TD realization is not limited in either time or bandwidth. The TD realization is
\begin{eqnarray} \label{TD of DD localized signal}
p_{\tau_0,\nu_0}(t) & = &  {\mathcal Z}_t^{-1}{\Big ( }  p_{_{\mbox{\scriptsize{dd}},\tau_0,\nu_0}} {\Big )} \nonumber \\
& & \hspace{-6mm} = \sqrt{\tau_{p}} \sum\limits_{n \in \mathbb{Z}} e^{j2\pi \nu_{0} n\tau_{p}} \, \delta(t-\tau_0-n\tau_{p}) \nonumber \\
& & \hspace{-6mm} = e^{j 2 \pi \nu_0 t} \,  \sqrt{\tau_{p}} \sum\limits_{n \in \mathbb{Z}} \delta(t-\tau_0-n\tau_{p}).
\end{eqnarray}We restrict the time and bandwidth of the pulsone by applying a factorizable pulse shaping filter
$w(\tau, \nu) = w_1(\tau) \, w_2(\nu)$, for example the Gaussian filter (\ref{filter eqn}). See \cite{zakotfs3}, Appendix \ref{app2} for a general treatment of factorizable filters. The filtered DD domain pulse centered at $(\tau_0, \nu_0) = (0,0)$ is given by
\begin{eqnarray}
\label{Filter DD siganl}
p_{_{\mbox{\scriptsize{dd}},0,0}}^w(\tau, \nu) & = & w (\tau,\nu) \ast_{\sigma} p_{_{\mbox{\scriptsize{dd}},0,0}}(\tau, \nu).
\end{eqnarray}It follows from (\ref{invzakt}) that the TD realization of the filtered DD domain pulse centered
at $(0,0)$ is given by
\begin{eqnarray}
\label{prevresult3}
    p_{_{\mbox{\scriptsize{Zak}}}}(t) & \Define & {\mathcal Z}_t^{-1}{\Big (}  p_{_{\mbox{\scriptsize{dd}},0, 0}}^w(\tau, \nu) {\Big )}  \nonumber \\
    & = & w_1(t) \, \star \,  {\Big (} W_2(t) \, \, p_{0,0}(t) {\Big )} \,\,\,,\,\,\, \mbox{\small{where}}\nonumber \\
\end{eqnarray}where $\star$ denotes linear convolution, and
\begin{eqnarray}
    W_2(t) & \Define & \int w_2(\nu) \, e^{j 2 \pi \nu t} \, d\nu,
\end{eqnarray}is the inverse Fourier transform of $w_2(\cdot)$. We limit the duration of the pulsone to $T$ by limiting the duration of $W_2(t)$ to $T$, that is by limiting the spread of the Fourier transform $w_2(\cdot)$ to $1/T$. We limit the bandwidth of the pulsone to $B$ by limiting the bandwidth of $w_1(t)$ to $B$, that is by limiting the spread of $w_1(t)$ to $1/B$. The delay spread of the pulse shaping filter $w(\tau, \nu)$ is roughly $1/B$ and the Doppler spread is roughly $1/T$. The number of non-overlapping DD domain pulses , each spread over an area $1/(B T)$,
inside a rectangle with width $\tau_p$ and height $\nu_p = 1/\tau_p$, is equal to the time-bandwidth product $BT$, rendering Zak-OTFS an orthogonal modulation that achieves the Nyquist rate. The pulse-shaping filter is normalized so that the total energy of the Zak-OTFS probe signal is $1$
\begin{eqnarray}
    \int \left\vert p_{_{\mbox{\scriptsize{Zak}}}}(t) \right\vert^2 \, dt & = & 1.
\end{eqnarray}
\textbf{Spread Pulsones:}
The peak to average power ratio (PAPR) of a pulsone is about $15$ dB since the TD waveform is a train of narrow pulses, exhibiting sharp peaks at the pulse locations. Transmission requires the use of highly linear power amplifiers which are typically power inefficient. However, it is possible to construct a spread pulsone for which the PAPR of the TD realization is about $6$ dB by applying a type of discrete spreading filter to a pulse in the DD domain (for details, see \cite{zakotfs3}). The effect of the spreading filter is to distribute energy equally across all $BT$ pulses in the DD domain. The effect of spreading in the TD is to produce a noise-like waveform that is much less peaky than the original pulsone.

\subsection{Ambiguity function}
The
ambiguity function of the Zak-OTFS probe $p_{_{\mbox{\scriptsize{Zak}}}}(t)$ is given by (\ref{appeqn2863}) (see top of next page). This expression follows from Appendix \ref{appn6} and the fact that $\ast_{\sigma}$ is an associative linear operator.
The delay spreads of the pulse shaping filters $w(\tau, \nu)$ and $w_{\mbox{\tiny{mf}}}(\tau, \nu)$ are roughly $1/B$ and the Doppler spreads are roughly $1/T$. The $(n,m)$-th term $A_{p,p,n,m}(\tau, \nu)$ is centered at the DD domain location $(n \tau_p, m \nu_p)$ and the spread in the delay and Doppler is similar to both $w(\tau, \nu)$ and $w_{\mbox{\tiny{mf}}}(\tau, \nu)$.
The high intensity regions are localized at the points of the period lattice $\Lambda_p$ given by
\begin{eqnarray}
    \Lambda_p & \Define & \{ (n \tau_p, m \nu_p) \, \vert \, n,m \in {\mathbb Z} \}.
\end{eqnarray}

\begin{figure*}
{\vspace{-9mm}
\begin{eqnarray}
\label{appeqn2863}
    A_{p,p}(\tau, \nu) &  =  & \sum\limits_{n \in {\mathbb Z}} \sum\limits_{m \in {\mathbb Z}}  w(\tau, \nu) \, \ast_{\sigma} \, \, {\Big (}  \delta(\tau - n \tau_p) \, \delta(\nu - m \nu_p) {\Big )} \, \ast_{\sigma} \, w_{\mbox{\tiny{mf}}}(\tau, \nu)  \,  = \,  \sum\limits_{n \in {\mathbb Z}} \sum\limits_{m \in {\mathbb Z}}  A_{p,p,n,m}(\tau, \nu), \nonumber \\
    A_{p,p,n,m}(\tau, \nu) & \Define & w(\tau, \nu) \, \ast_{\sigma} \, {\Big (} e^{j 2 \pi m \nu_p \tau} \, w_{\mbox{\tiny{mf}}}(\tau - n \tau_p, \nu - m \nu_p) {\Big )}.
\end{eqnarray}
\vspace{-4mm}
\begin{eqnarray*}
    \hline
\end{eqnarray*}
\normalsize}
\end{figure*}
Fig.~\ref{otfs plusone} illustrates the intensity
of the self-ambiguity function $A_{p,p}(\tau, \nu)$
with Gaussian pulse shaping.
\begin{figure}
      \includegraphics[scale=0.13]{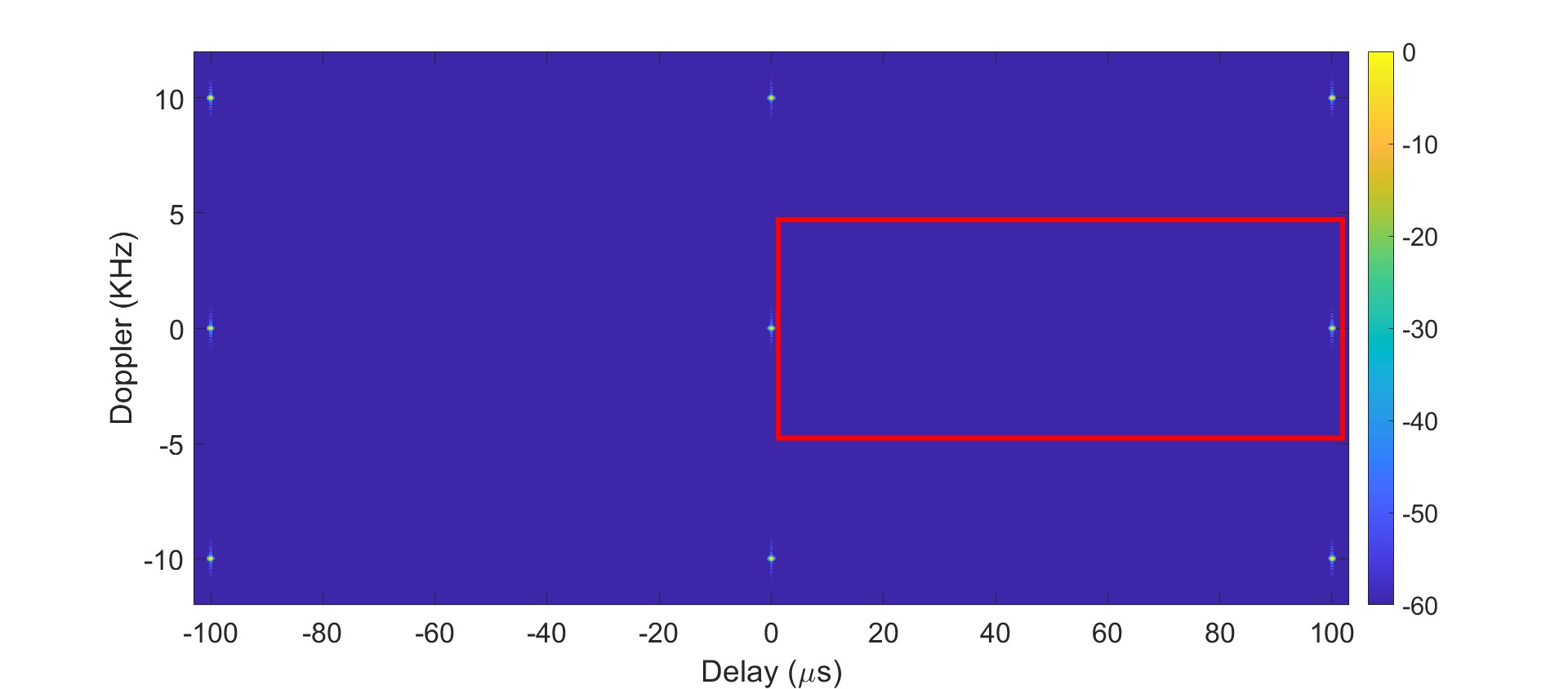}
  \caption{Heat map illustrating the intensity
  of the self-ambiguity function $A_{p,p}(\tau, \nu)$
  for the Zak-OTFS probe. The bandwidth is $B = 4$ MHz,
  the time duration $T = 20$ ms, the delay period $\tau_p = 100 \mu s$ and the Doppler period $\nu_p = 10$ KHz. The intensity pattern is characteristic of Gaussian pulse shaping. As an example, if the delay and Doppler shifts of the targets satisfy $\tau_{min} = 0$, $\tau_{max}=90 \mu s$, and $-\nu_{min} =  \nu_{max} = 4 $ KHz, then the crystallization condition is satisfied, and target locations can be estimated within the rectangle with the red border.}
  \label{otfs plusone}
     \end{figure}

\textbf{Spread Pulsones:} The self-ambiguity function of the Zak-OTFS probe is supported on the period lattice $\Lambda_p$. When we spread the Zak-OTFS probe by applying a discrete chirp filter in the DD domain, we obtain a spread pulsone with a self-ambiguity function that is supported on a rotated lattice $\Lambda$ (for more information see \cite{zakotfs3}).
This is similar in spirit to the way radar engineers construct waveform libraries for tracking applications consisting of an initial waveform that has been linearly
frequency modulated or chirped at various rates (see \cite{HSM} for more details).

\subsection{Noise-free target detection}
We consider a radar scene $h(\tau, \nu)$ with $P$ targets, where the return from the $k$-th target is delayed by $\tau_k$, shifted in Doppler by $\nu_k$, and scaled by a complex number $h_k$.
\begin{eqnarray}
    h(\tau, \nu) & = & \sum\limits_{i=1}^P h_i \, \delta(\tau - \tau_i) \, \delta(\nu - \nu_i). 
\end{eqnarray}It follows from (\ref{sim_cross_ambiguity_eqn}) that the cross-ambiguity between the radar response $y(t)$ and the Zak-OTFS probe $p_{_{\mbox{\scriptsize{Zak}}}}(t)$ is given by
\begin{eqnarray}
    A_{y,p}(\tau, \nu) & = & h(\tau, \nu) \, \ast_{\sigma} \, A_{p,p}(\tau, \nu) \nonumber \\
    & = & \sum\limits_{i=1}^P h_i \, {\Big (} \delta(\tau - \tau_i) \, \delta(\nu - \nu_i) {\Big )} \, \ast_{\sigma} \, A_{p,p}(\tau, \nu)\, \nonumber \\
    & = & \sum\limits_{i=1}^P h_i \, A_{p,p}(\tau - \tau_i, \nu - \nu_i) \, e^{j 2 \pi \nu_i (\tau - \tau_i)}.
\end{eqnarray}
We substitute for $A_{p,p}(\tau, \nu)$ using (\ref{appeqn2863}) to obtain
\begin{eqnarray}
\label{eqn8924316}
    A_{y,p}(\tau, \nu) & \hspace{-3mm} = & \hspace{-3mm} \sum\limits_{n,m \in {\mathbb Z}} \hspace{-2mm}  B_{n,m}(\tau, \nu) \,\,,\,\, \nonumber \\
    B_{n,m}(\tau, \nu) & \hspace{-3mm} \Define & \hspace{-3mm} \sum\limits_{i=1}^P \hspace{-1mm} h_i \, e^{j 2 \pi \nu_i (\tau - \tau_i)} A_{p,p,n,m}(\tau - \tau_i, \nu - \nu_i). \nonumber \\
\end{eqnarray}Let $\tau_{max}, \tau_{min}$ respectively denote the maximum and minimum path delay shift, and
$\nu_{max}, \nu_{min}$ denote the maximum and minimum path Doppler shift respectively. In (\ref{eqn8924316}), the term $A_{p,p,n,m}(\tau - \tau_i, \nu - \nu_i)$  has a peak at $(\tau_i + n \tau_p, \nu_i + m \nu_p)$. Therefore the support set for $B_{n,m}(\tau, \nu)$ lies within the DD rectangle
\begin{eqnarray}
    {\mathcal D}_{n,m} & \Define & {\Big \{} (\tau, \nu) \, \vert \, \tau_{min} + n \tau_p \leq \tau \leq  \tau_{max} + n \tau_p\,,\, \nonumber \\
    & & \nu_{min} + m \nu_p \leq \nu \leq \nu_{max} + m \nu_p  {\Big \}}.
\end{eqnarray}
We choose the delay period $\tau_p$ to be greater than the delay spread $(\tau_{max} - \tau_{min})$ and we choose the Doppler period to be greater than the Doppler spread $(\nu_{max} - \nu_{min})$. This is the \emph{crystallization condition} that prevents aliasing in Zak-OTFS communication systems (see \cite{zakotfs1}, \cite{zakotfs2} for more details). In this radar application it implies that for all $(n_1, m_1) \ne (n_2, m_2)$
\begin{eqnarray}
    {\mathcal D}_{n_1, m_1} \, \bigcap \, {\mathcal D}_{n_2, m_2} & = & \phi
\end{eqnarray}To see this, suppose $n_1 < n_2$. If there were a point $(a,b)$ in the intersection then
\begin{eqnarray}
    \tau_{min} + n_2 \tau_p & < & a \, < \, \tau_{max} + n_1 \tau_p,
\end{eqnarray}which implies $\tau_p < \tau_{max} - \tau_{min}$, contradicting our choice of $\tau_p$.
Since the support rectangles do not overlap when the crystallization condition is satisfied, there are exactly $P$ target peaks in each rectangle ${\mathcal D}_{m,n}$, and for each target, we can estimate the delay and Doppler shift from the difference between the location of the peak and $(n \tau_p, m \nu_p)$.

For example, suppose the bandwidth is $1$ MHz, the time duration $T = 10$ ms, the delay period $\tau_p = 100 \mu s$ and the Doppler period $\nu_p = 10$ KHz. If $\tau_{min} = 0$, $\tau_{max} = 90 \mu s$, and $-\nu_{min} = \nu_{max} = 4$ KHz, then the crystallization condition is satisfied, and target locations can be estimated within the rectangle with the red border shown in Fig.~\ref{otfs plusone}.

If the crystallization condition were not satisfied, then target detection would be ambiguous, since overlapping support rectangles result in false peaks
within each rectangle $D_{m,n}$. This is the counterpart
of DD domain aliasing in Zak-OTFS communication systems (see \cite{zakotfs1, zakotfs2} for more details).

\textbf{Radar Resolution:} We now suppose that the crystallization condition holds, and we show that Zak-OTFS probes provide radar resolution of $1/B$ along the delay axis and $1/T$ along the Doppler axis. We look to separate two targets at $(\tau_1, \nu_1)$ and $(\tau_2, \nu_2)$ by separating the corresponding cross-ambiguity terms within $B_{0,0}(\tau, \nu)$. Within $B_{0,0}(\tau, \nu)$, the
term $A_{p,p,0,0}(\tau - \tau_1, \nu - \nu_1)$ corresponding to the first target is centered at $(\tau_1, \nu_1)$, and the spread in delay and Doppler is roughly the same as the pulse shaping filter $w(\tau, \nu)$, which is $1/B$ in delay and $1/T$ in Doppler.
The same reasoning applies to the second target. Hence the two targets will appear as separate peaks in the cross-ambiguity function only if $\vert \tau_1 - \tau_2 \vert > \frac{1}{B}$ or $\vert \nu_1 - \nu_2 \vert > \frac{1}{T}$.

\textbf{Examples:}
Suppose the bandwidth $B = 4$ MHz, the time duration $T=20$ ms, the delay period $\tau_p = 100 \mu s$, and the
Doppler period $\nu_p = 10$ KHz. The delay resolution
is $1/B = 0.25 \mu s$ and the Doppler resolution is $1/T = 50$ Hz. Consider three targets located at
$(0.6 \mu s, -220)$ Hz, $(0.95 \mu s, -220)$ Hz and $(0.6 \mu s, -290)$ Hz. For the purpose of illustration only, we consider
no additional reflectors and path-loss to be same as in free space. Specifically, we consider $\vert h_i \vert = 10^{-7}/\tau_i$. Fig.~\ref{otfs plusone2}
illustrates the intensity of the cross-ambiguity function $A_{y,p}(\tau, \nu)$. Targets at $(0.6 \mu s, -220)$ Hz and $(0.95 \mu s, -220)$ Hz have the same Doppler location but are separated along delay by $0.35 \, \mu s$ which is greater than the delay resolution of $0.25 \mu s$ and are hence separable. Similarly, the targets at 
$(0.6 \mu s, -220)$ Hz and $(0.6 \mu s, -290)$ have the same delay location but are separated along Doppler by $70$ Hz which is more than the Doppler resolution of $50$ Hz and are hence separable.

\begin{figure}
    \includegraphics[width=1.1\linewidth]{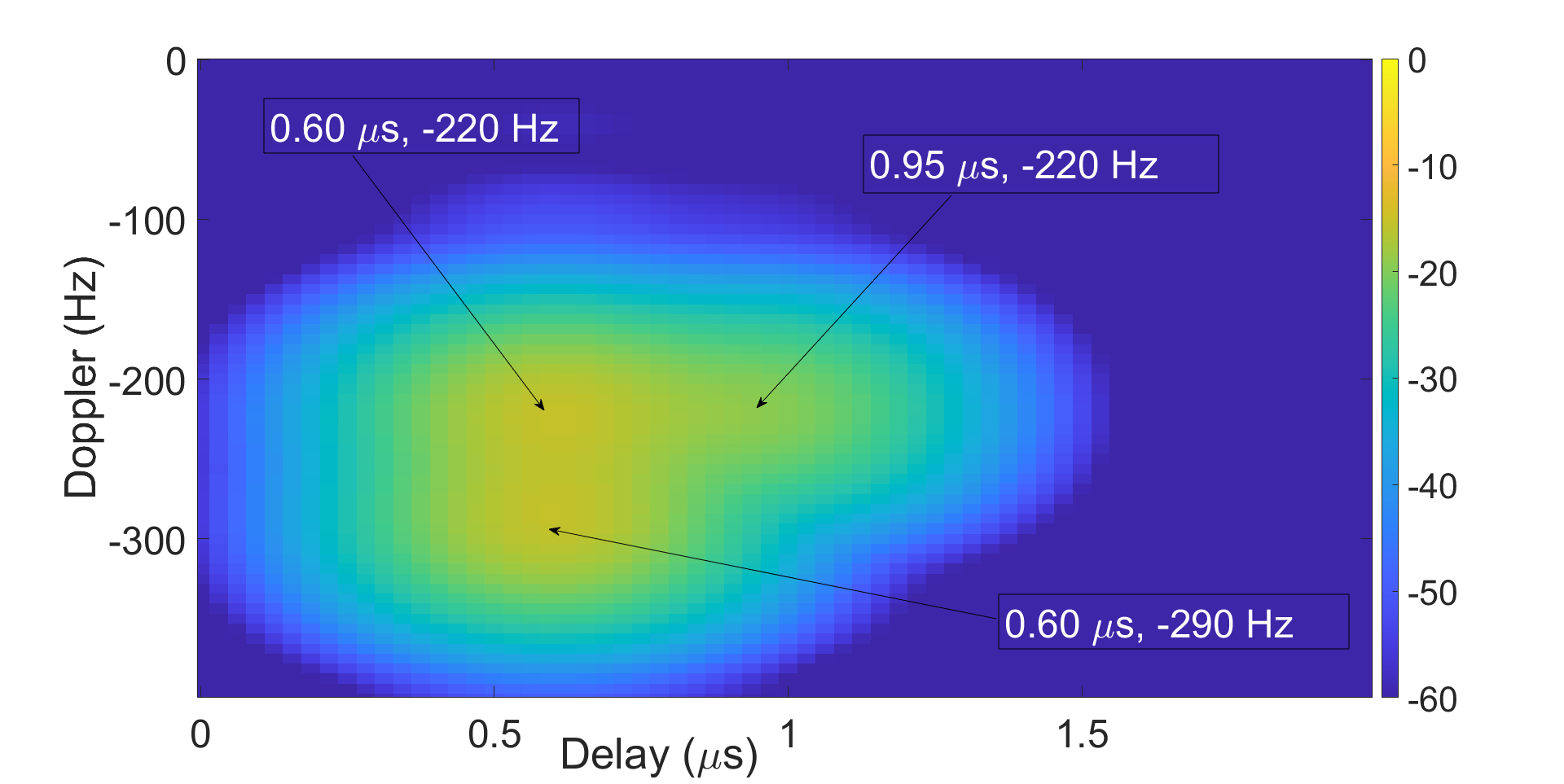}
  \caption{Heat map illustrating the intensity of the cross-ambiguity function $A_{y, p}(\tau, \nu)$ for the Zak-OTFS probe.
  The bandwidth is $B = 4 $ MHz, the time-duration
  $T = 20$ ms, the delay period $\tau_p = 100 \mu s$, and the Doppler period $\nu_p = 10$ KHz.
  There are three targets located at $(0.6 \mu s, -220)$ Hz, $(0.95 \mu s, -220)$ Hz and $(0.6 \mu s, -290)$ Hz. Noise-free radar processing. Three peaks corresponding to the three targets are separable/resolvable since any two targets are well separated along either delay or Doppler axis.}
  \label{otfs plusone2}
     \end{figure}

Fig.~\ref{otfs plusone3} provides an example where it is not possible to resolve targets because the minimum delay
spacing is less than the delay resolution and the minimum Doppler spacing is less than the Doppler resolution.

\begin{figure}
    \includegraphics[width=1.1\linewidth]{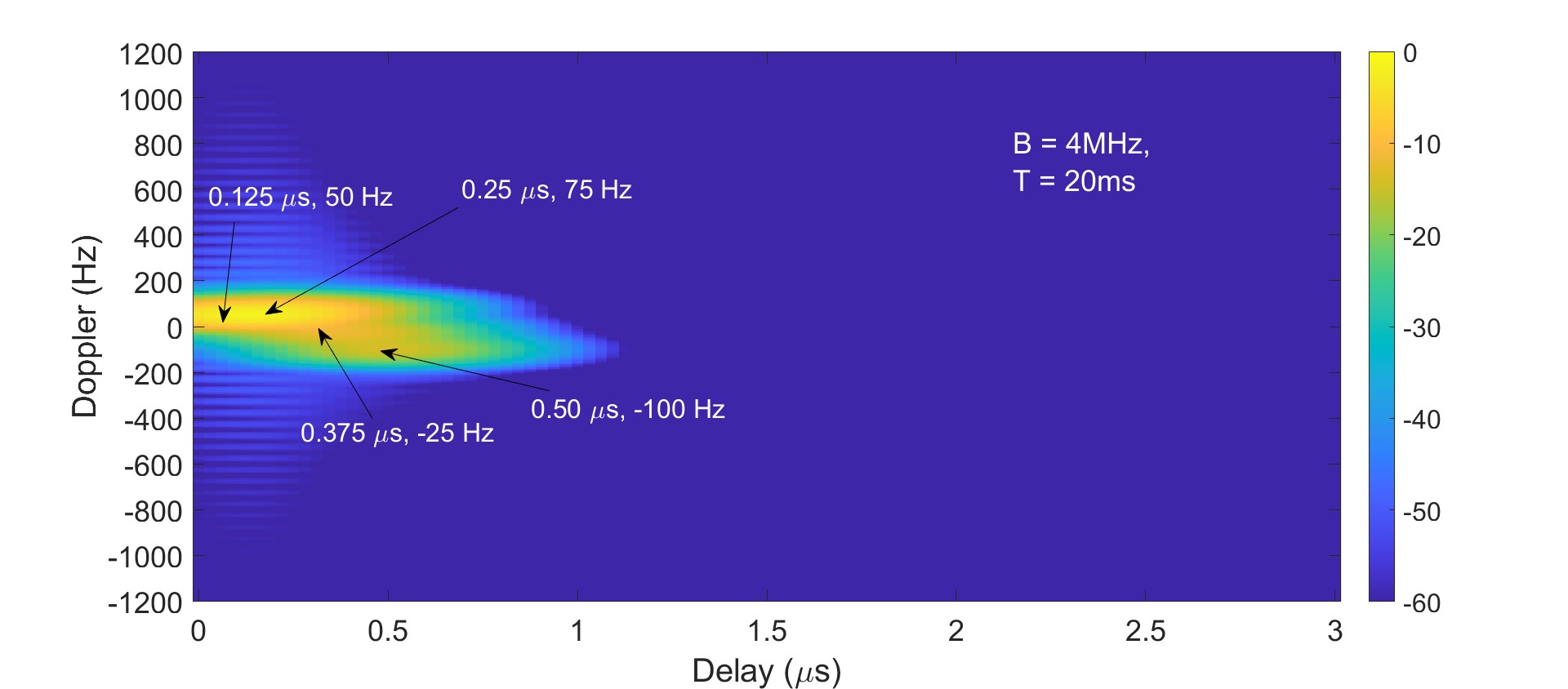}
  \caption{Heat map illustrating the intensity of the cross-ambiguity function $A_{y, p}(\tau, \nu)$ for the Zak-OTFS probe.
  The bandwidth is $B = 4 $ MHz, the time-duration
  $T = 20$ ms, the delay period $\tau_p = 100 \mu s$, and the Doppler period $\nu_p = 10$ KHz.
  There are four targets located at $(0.125 \mu s, 50)$ Hz, $(0.25 \mu s, 75)$ Hz, $(0.375 \mu s, -25)$ Hz and
$(0.5 \mu s, -100)$ Hz. Noise-free radar processing.
It is not possible to resolve the four peaks corresponding to the four targets since the targets are not well separated in both delay and Doppler.}
  \label{otfs plusone3}
     \end{figure}
\textbf{Number of resolvable targets:} The delay resolution $\Delta \tau \propto 1/B$ and the Doppler resolution $\Delta \nu \propto 1/T$. For unambiguous target detection we restrict to DD locations $(\tau, \nu)$ within the fundamental period of the period lattice $\Lambda_p$, i.e., $0 \leq \tau < \tau_p$ and $0 \leq \nu < \nu_p$.
Since the fundamental period has unit area ($\tau_p \, \nu_p = 1$), the number of resolvable targets is $\frac{1}{\Delta \tau} \, \frac{1}{\Delta \nu} = O(BT)$.

\textbf{Practical Implementation and Complexity:}
For practical implementation, the Zak-transform of the received TD signal $y(t)$ (i.e., $y_{\mbox{\scriptsize{dd}}}(\tau, \nu)$) is computed on a finite number of points of a DD domain lattice (which is finer than the delay and Doppler resolution). Let positive integer constants $P$ and $Q$ denote the oversampling factors along the delay and Doppler domain. To be precise, $y_{\mbox{\scriptsize{dd}}}(\tau, \nu)$ is computed for $(\tau,\nu)$ in the set
\begin{eqnarray}
    {\mathcal S} & \Define & {\Big \{} (\tau, \nu) \, {\Big \vert}  \, \tau = \frac{k}{P B} \,,\, \nu = \frac{l}{Q T} \,,\, k,l \in {\mathbb Z}\nonumber \\
    & & \hspace{-9mm} 0 \leq k \leq PM -1  
    \,,\,  0 \leq l \leq QN -1  {\Big \}}, \,\, \mbox{where}
   \end{eqnarray}
\begin{eqnarray}
\label{mndef}
     M & \Define & B \tau_p \,,\, N  \Define  T \nu_p
\end{eqnarray}are integers. The cardinality of ${\mathcal S}$ is $PQMN$. Similarly, the Zak-transform of the transmitted Zak-OTFS probe waveform is also pre-computed at $(\tau, \nu) \in {\mathcal S}$ and stored. From (\ref{zaktransformeqn}), the Zak-transform of $y_{\mbox{\scriptsize{dd}}}(\tau, \nu)$ with $(\tau, \nu) \in {\mathcal S}$ is given by
\begin{eqnarray}
\label{sampledzak}
  y_{\mbox{\scriptsize{dd}}}\left(\frac{k}{PB}, \frac{l}{QT}\right) & \hspace{-2.5mm} = & \hspace{-2.5mm} \sqrt{\tau_p} \sum\limits_{n = - \lceil \frac{N}{2} \rceil }^{N - \lceil \frac{N}{2} \rceil } \hspace{-1mm} y\left( \frac{k + nPM}{PB} \right) \, e^{-j 2 \pi \frac{ n l}{QN}}, \nonumber \\
\end{eqnarray}$k = 0,1, \cdots,PM-1 \,\,$,$\,\, l = 0,1,\cdots, QN -1$.
 In (\ref{sampledzak}), the lower and upper indices
of the summation variable $n$ follows from the fact that the transmitted probe waveform has almost all of its energy limited to the time-interval
$\left[  -\frac{T}{2} \,,\ \frac{T}{2}\right] $. For each $k \in \{0, 1, \cdots,  PM-1 \}$, $y_{\mbox{\scriptsize{dd}}}\left(\frac{k}{PB}, \frac{l}{QT}\right)$ can be computed for all $l \in \{0, 1, \cdots,  QN-1 \}$  as a $QN$-point Fast Fourier Transform (FFT) which has complexity $O(QN \log(QN))$ (i.e., $O(N \log (N))$ since $Q$ is constant). Hence the complexity of computing $y_{\mbox{\scriptsize{dd}}}(\tau, \nu)$ for all $(\tau, \nu) \in {\mathcal S}$ is $O(MN \log(N))$. Since $MN = BT$ (see (\ref{mndef})), the complexity is $O(BT \log(BT))$. 

Next, we compute the cross-ambiguity $A_{y,x}(\tau, \nu)$ in (\ref{crossambigdd}) for $(\tau, \nu) \in  [\tau_{min}  \,,\,  \tau_{max}]\, \times \, [-\nu_{max}  \,,\,  \nu_{max}]$. For implementation purposes, both $y_{\mbox{\scriptsize{dd}}}(\tau, \nu) $ and $x_{\mbox{\scriptsize{dd}}}(\tau, \nu)$ are represented by their discrete DD domain representation over the set ${\mathcal S}$ and therefore it suffices to compute $A_{y,x}(\tau, \nu)$ for $(\tau, \nu)$ belonging to the set

{\small
\vspace{-4mm}
\begin{eqnarray}
    {\mathcal S}_c & \Define & {\Big \{} (\tau, \nu) \, {\Big \vert}  \, \tau = \frac{k}{P B} \,,\, \nu = \frac{l}{Q T} \,,\, k,l \in {\mathbb Z}\nonumber \\
    & & \hspace{-13mm} P\lfloor B \tau_{min} \rfloor \leq k \leq P\lceil B \tau_{max} \rceil   
    \,,\,  - Q\lceil T \nu_{max} \rceil  \leq l \leq Q\lceil T \nu_{max} \rceil   {\Big \}}. \nonumber \\
\end{eqnarray}\normalsize}
The cardinality of ${\mathcal S}_c$ is $O(BT)$ since we assume underspread channels, i.e., $2 (\tau_{max} - \tau_{min}) \nu_{max} < 1$ (product of channel delay and Doppler spread is less than one).
The cross-ambiguity integral in the R.H.S. of (\ref{crossambigdd2}) (see top of next page), when computed at $(\tau, \nu) \in {\mathcal S}_c$ is closely approximated by its Riemann sum given by the R.H.S. in the second line of (\ref{crossambigdd2}).
For the Zak-OTFS probe waveform, its DD representation $x_{\mbox{\scriptsize{dd}}}(\tau, \nu)$ is sparse and localized around the lattice points of the period lattice $\Lambda_p$ (see (\ref{Filter DD siganl})).
Note that the cross-ambiguity integration variables are limited to one period of the lattice $\Lambda_p$, i.e., $0 < \tau' < \tau_p$ and $0 < \nu' < \nu_p$.
Within one lattice period, the delay and Doppler spread of $x_{\mbox{\scriptsize{dd}}}(\tau, \nu)$ is the same as the delay and Doppler spread of the pulse shaping filter $w(\tau, \nu)$, which is $O(1/B)$ and $O(1/T)$ respectively. Hence the number of significant terms in the summation in the R.H.S. of (\ref{crossambigdd2}) is constant and does not depend on $B$ and $T$ (for the Gaussian pulse, it is roughly $5P \times 5Q = 25PQ$, i.e., five resolutions along delay and five along Doppler).
Therefore, the complexity of computing $A_{y,x}(\tau, \nu)$ for all $(\tau, \nu) \in {\mathcal S}_c$ is proportional to the cardinality of ${\mathcal S}_c$ and is therefore $O(BT)$. The overall complexity of computing the discrete DD representation of the received waveform and then computing the cross-ambiguity function is therefore $O(BT \, \log(BT))$.

\begin{figure*}
\vspace{-9mm}
{\small
\begin{eqnarray}
\label{crossambigdd2}
    A_{y,x}\left( \frac{k}{PB}, \frac{l}{QT} \right) &  =  & \int\limits_{0}^{\tau_p} \hspace{-1mm} \int\limits_{0}^{\nu_p} \hspace{-1mm} y_{\mbox{\scriptsize{dd}}}(\tau', \nu') \, x_{\mbox{\scriptsize{dd}}}^*(\tau' - k/(PB), \nu' - l/(QT)) \, e^{-j 2 \pi \frac{l}{QT} \left(\tau' - \frac{k}{PB} \right)} \, d\tau' \, d\nu' \nonumber \\
    & \approx & \frac{1}{PQ M N} \sum\limits_{k'=0}^{PM-1} \sum\limits_{l'=0}^{QN -1} y_{\mbox{\scriptsize{dd}}}\left(\frac{k'}{PB}, \frac{l'}{QT} \right) \, x_{\mbox{\scriptsize{dd}}}^*\left(\frac{(k' - k)}{PB}, \frac{(l' - l)}{QT} \right) \, e^{- j 2 \pi \frac{l (k' - k)}{PQ MN}}
\end{eqnarray}\normalsize}
\vspace{-3mm}
\begin{eqnarray*}
    \hline
\end{eqnarray*}
\end{figure*}

\textbf{Spread Pulsones:} The self-ambiguity function
of a spread pulsone is supported on a lattice $\Lambda$
obtained by rotating the period lattice $\Lambda_p$ \cite{zakotfs3}. The crystallization condition
for the spread pulsone is that the translates of the support rectangle ${\mathcal D}_{0,0}$ by lattice points
in $\Lambda$ should not overlap. It is possible to resolve targets when this condition is satisfied.

\section{Numerical Simulations}
\label{simsec}
\begin{figure}
\hspace{-5mm]}
    \includegraphics[width=9.4cm,height=5.5cm]{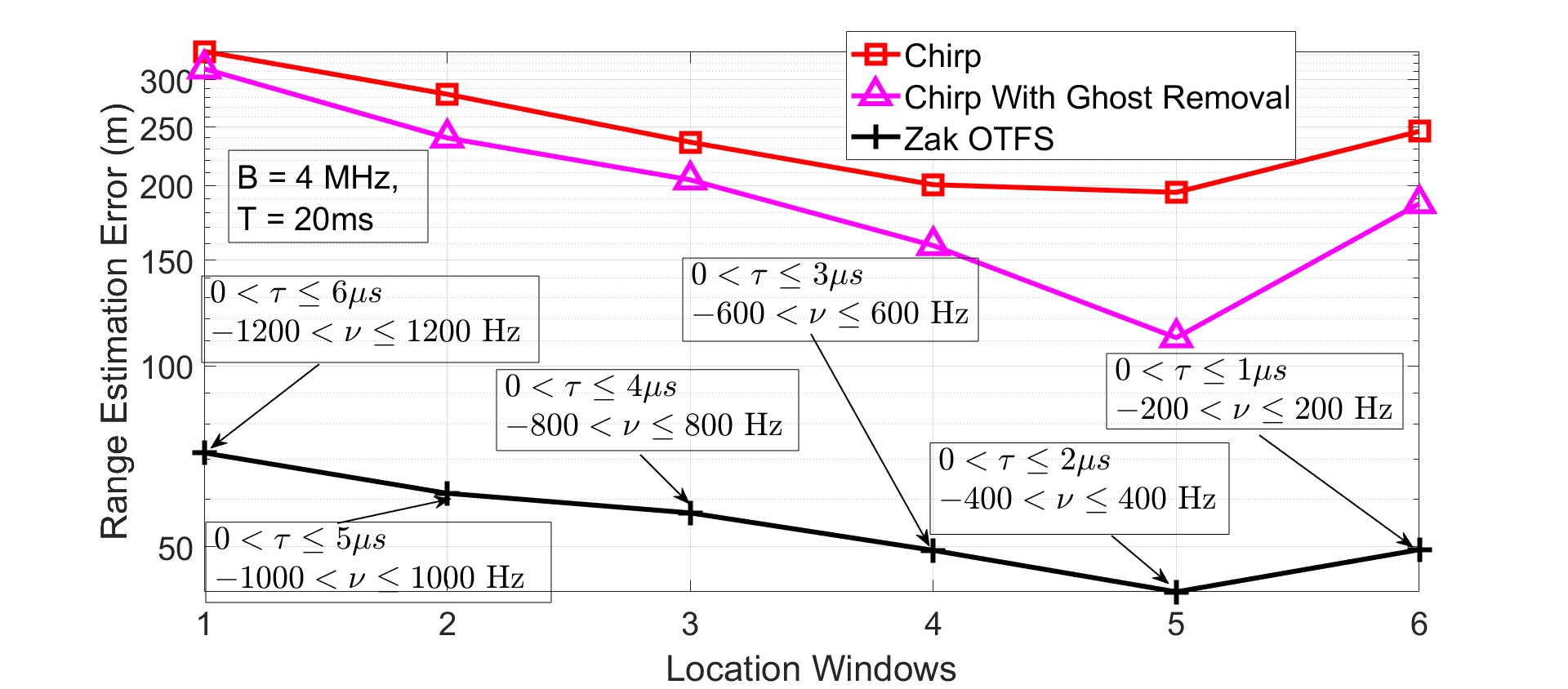}
  \caption{RMS range estimation error for four targets
  uniformly distributed in six rectangles that shrink from left to right. Bandwidth $B = 4$ MHz and a time duration $T=20$ ms. Zak-OTFS waveform with $\tau_p = 100 \, \mu s$ and $\nu_p = 10$ KHz. Gaussian pulse-shaping filter (\ref{filter eqn}) with $\alpha = \beta = 1.584$. No noise, i.e., infinite SNR.}
  \label{rangeerror_nonoise}
     \end{figure}
All simulations assume a bandwidth $B = 4$ MHz and a time duration $T=20$ ms. We first consider noise-free estimation of the range and velocity of multiple targets using a Zak-OTFS waveform and two different chirp waveforms. The delay period of the Zak-OTFS waveform is $\tau_p = 100 \, \mu s$, and the Doppler period $\nu_p = 10$ KHz. The first chirp waveform consists of an up-chirp with slope $2B/T$ transmitted in the first $T/2$ seconds followed by a down-chirp with slope $-2B/T$ transmitted in the subsequent $T/2$ seconds. The second chirp waveform consists of two pairs of up-chirp/down-chirp signals. The first pair has slopes $2B/T$ and $-2B/T$ transmitted in the first $T/2$ seconds, the second pair has slopes $4B/T$ and $-4B/T$ and are transmitted in the subsequent $T/2$ seconds. These are the chirp waveforms described in Section \ref{chirpsec}. 

In each case we detect targets using the cross-ambiguity between the return signal and the probe waveform. The peaks in the cross-ambiguity function point to target locations as described in Sections \ref{chirpsec} and \ref{sec4}. In each case we employ a Gaussian pulse-shaping filter (\ref{filter eqn}) with $\alpha = \beta = 1.584$.

Fig.~\ref{rangeerror_nonoise} illustrates root mean squared (RMS) range estimation error for four targets uniformly distributed in six rectangles $\Omega_i = \left[ 0 , (7 -i)\right] \mu s \, \times \left[ -200(7-i) , 200(7 -i)\right]$ Hz, $i=1,2,\cdots, 6$ (x-axis in Fig.~\ref{rangeerror_nonoise} is $i=1,2,\cdots, 6$). Target spacing decreases as $i$ increases. Estimated range is half the product of the target delay and the speed of light. Estimated velocity is half the product of the target Doppler shift and the RF carrier wavelength (corresponding to a carrier frequency of $1$ GHz).
For all waveforms, the RMS error decreases from $\Omega_1$ to $\Omega_5$ as the rectangle shrinks, and the RMS error increases from $\Omega_5$ to $\Omega_6$ as the targets become too close to separate. Fig.~\ref{rangeerror_nonoise} confirms that using two pairs of up-chirp/down-chirp signals to remove \emph{ghosts} (referred to as ``Chirp with ghost removal") results in lower RMS error than the single pair of up-chirp/down-chirp signals \cite{Bajwa}. However, for the chirp waveform with two-pairs of up-chirp/down-chirp, the duration of each individual chirp signal is $T/4$ as compared to $T/2$ for the chirp waveform with only a single up-chirp/down-chirp. This limits the resolution of the chirp waveform with two pairs of up-chirp/down-chirp resulting in higher RMS error when compared to Zak-OTFS. Fig.~\ref{velerror_nonoise} shows that the advantages of Zak-OTFS extend to velocity estimation.

We next consider the accuracy of estimating four targets uniformly distributed in the rectangle $[0 \,,\, 3] \mu s \, \times \, [-600 \,,\, 600]$ Hz, as a function of SNR. Fig.~\ref{rangeerror_withnoise} illustrates that at low SNR, the RMS error for range estimation is almost identical for chirp and Zak-OTFS waveforms, but with increasing SNR, the RMS performance of Zak-OTFS is significantly better.  Fig.~\ref{velerror_withnoise} illustrates similar result for RMS velocity estimation error.
     
\begin{figure}
\hspace{-5mm]}
    \includegraphics[width=9.4cm,height=5.5cm]{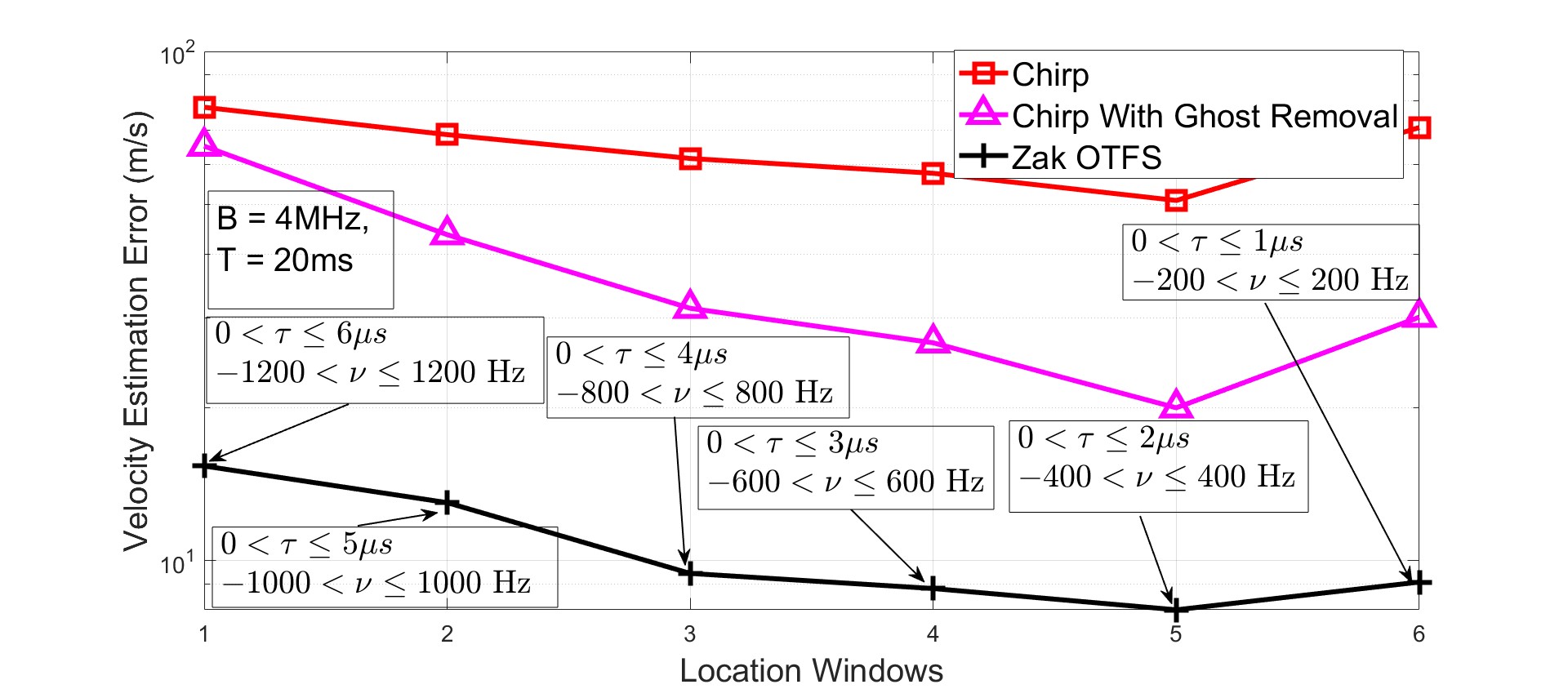}
  \caption{RMS velocity estimation error for four targets
  uniformly distributed in six rectangles that shrink from left to right. Bandwidth $B = 4$ MHz and a time duration $T=20$ ms. Zak-OTFS waveform with $\tau_p = 100 \, \mu s$ and $\nu_p = 10$ KHz. Gaussian pulse-shaping filter (\ref{filter eqn}) with $\alpha = \beta = 1.584$. No noise, i.e., infinite SNR.}
  \label{velerror_nonoise}
     \end{figure}

\begin{figure}
\hspace{-5mm]}
    \includegraphics[width=9.4cm,height=5.5cm]{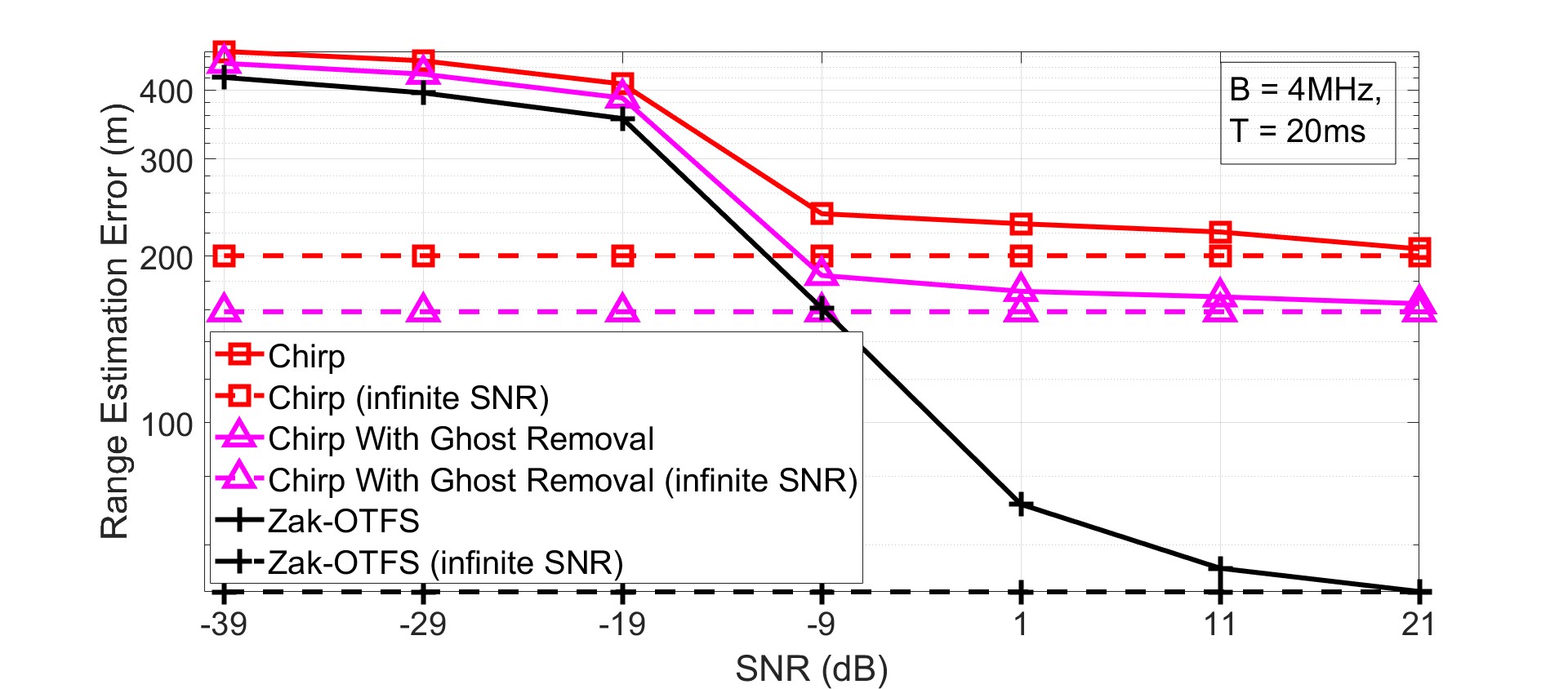}
  \caption{RMS range estimation error for chirp and Zak-OTFS waveforms as a function of increasing SNR. Four targets uniformly distributed in the rectangle $[0 \,,\, 3] \mu s\, \times \, [-600 \,,\, 600]$ Hz. Bandwidth $B = 4$ MHz and a time duration $T = 20$ ms. Zak-OTFS waveform with $\tau_p = 100 \, \mu s$ and $\nu_p = 10$ KHz. Gaussian pulse shaping filter (\ref{filter eqn}) with $\alpha = \beta = 1.584$.}
  \label{rangeerror_withnoise}
     \end{figure}

     \begin{figure}
\hspace{-5mm]}
    \includegraphics[width=9.4cm,height=5.5cm]{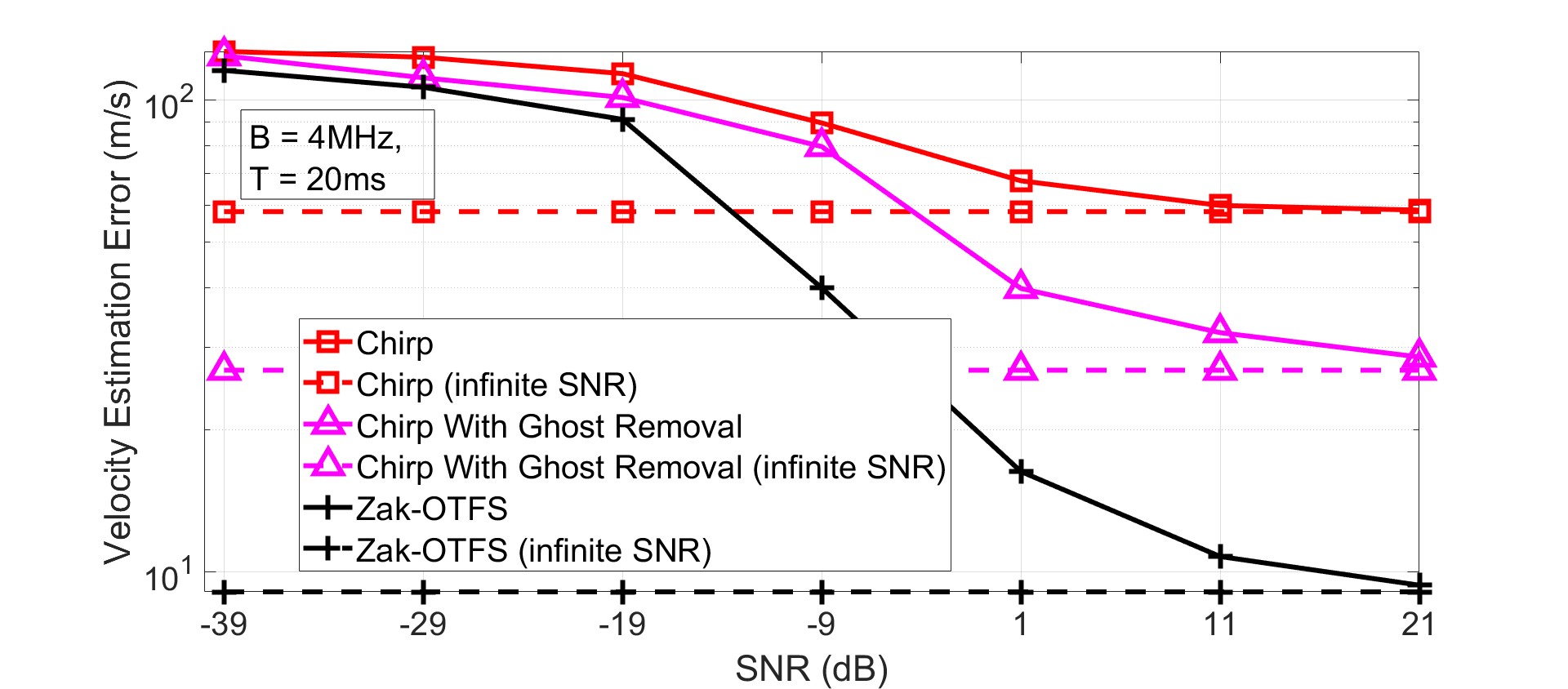}
  \caption{RMS velocity estimation error for chirp and Zak-OTFS waveforms as a function of increasing SNR. Four targets uniformly distributed in the rectangle $[0 \,,\, 3] \mu s\, \times \, [-600 \,,\, 600]$ Hz. Bandwidth $B = 4$ MHz and a time duration $T = 20$ ms. Zak-OTFS waveform with $\tau_p = 100 \, \mu s$ and $\nu_p = 10$ KHz. Gaussian pulse shaping filter (\ref{filter eqn}) with $\alpha = \beta = 1.584$.}
  \label{velerror_withnoise}
     \end{figure}

\section{Conclusion}
LTV systems model radar scenes where each reflector/ target applies a delay, Doppler shift and complex amplitude scaling to a transmitted waveform. We described how the receiver uses the transmitted signal as a reference, how target locations correspond to regions of high intensity in the cross-ambiguity between the received signal and the transmitted waveform. We described how the self-ambiguity function of the transmitted waveform limits resolution of the delay and Doppler shifts of multiple targets in close proximity. In other words, we described how target resolution is governed by the geometry of the self-ambiguity function of the transmitted waveform. We compared traditional chirp waveforms, for which the self-ambiguity function is a line, with the Zak-OTFS carrier waveform, for which the self-ambiguity function is a lattice. We provided numerical simulations for clusters of targets illustrating that the lattice geometry affords superior target resolution.

\appendices

\section{Derivation of Eq. (\ref{MLEeqn})}
\label{app1}
With only one reflecting target
the received signal is
\begin{eqnarray}
y(t) & = & h_1 \, x(t - \tau_1) \, e^{j 2 \pi \nu_1 (t - \tau_1)} \, + \, n(t).
\end{eqnarray}
Since $n(t)$ is AWGN, the MLE estimate is given by
\begin{eqnarray}
\label{app1_eqn1}
 (\hat{\tau_1},\hat{\nu_1}) & \hspace{-2mm} \overset{\Delta}{=}& \hspace{-2mm} \underset{\tau,\nu}{argmin} \, \min_{h} \int \left\vert y(t) - h \,  x(t - \tau) \, e^{j 2 \pi \nu (t - \tau)} \right\vert^2 \, dt. \nonumber \\
\end{eqnarray}We firstly analyze the inner minimization w.r.t. $h$.
Note that
\begin{eqnarray}
\label{app1_eqn2}
    \int \left\vert y(t) - h x(t - \tau) \, e^{j 2 \pi \nu (t - \tau)} \right\vert^2 \, dt & & \nonumber \\
    & & \hspace{-59mm} =  \hspace{-1mm} \int \hspace{-1mm} \left\vert y(t) \right\vert^2 \, dt \, + \, \vert h \vert^2 \hspace{-1mm} \int \hspace{-1mm} \vert x(t) \vert^2 dt  - 2 \Re\left[ h^* A_{y,x}(\tau, \nu)\right]. 
\end{eqnarray} Since the received signal energy $\int \vert y(t) \vert^2 dt$ does not depend on $(h, \tau, \nu)$, it suffices to consider the inner minimization in (\ref{app1_eqn1}) to be
\begin{eqnarray}
    \min_{\vert h \vert} \left( \vert h \vert^2 E_T \, - \, \vert h \vert \, \vert A_{y,x}(\tau, \nu) \vert \right)  & = & - \frac{\left\vert A_{y,x}(\tau, \nu) \right\vert^2}{4 E_T}
\end{eqnarray}where $E_T \overset{\Delta}{=} \int \hspace{-1mm} \vert x(t) \vert^2 dt$ is the energy of the transmitted radar signal. Using this in (\ref{app1_eqn1}), the MLE is given by
\begin{eqnarray}
(\hat{\tau_1},\hat{\nu_1}) & \hspace{-2mm} \overset{\Delta}{=}& \hspace{-2mm} \underset{\tau,\nu}{argmin}  \frac{-\left\vert A_{y,x}(\tau, \nu) \right\vert^2}{4 E_T}.
\end{eqnarray}Since $E_T$ is constant, this is same as (\ref{MLEeqn}).

\section{Derivation of Eq.(\ref{sim_cross_ambiguity_eqn})}
\label{app2}
Using the expression for the received signal $y(t)$ from (\ref{otfsrxsigneqn}) in (\ref{cross_ambiguity_eqn}) we get (\ref{app2_eqn1}) (see top of next page).
Substituting $t' = (t - \tau')$ in the inner integral of the first term in the second step of (\ref{app2_eqn1}) gives the expression in the
third step. The inner integral in the third step is clearly $A_{x,x}(\tau - \tau', \nu - \nu')$ where $A_{x,x}(\tau, \nu)$ is given by
(\ref{self_ambiguity_eqn}).
\begin{figure*}
\vspace{-9mm}
\begin{eqnarray}
\label{app2_eqn1}
    A_{y,x}(\tau, \nu) & \hspace{-2mm} = &  \hspace{-2mm} \iiint h(\tau', \nu') x(t - \tau') e^{j 2 \pi \nu' (t - \tau')} x^*(t - \tau) e^{-j 2 \pi \nu(t - \tau)} \, dt \, d\tau'  \, d\nu' \,\, + \, \int n(t) \, x^*(t - \tau) \, e^{-j 2 \pi \nu (t - \tau)} \, dt \nonumber \\
    & & \hspace{-21mm} =  \iint h(\tau', \nu') \, e^{j 2 \pi \nu' (\tau - \tau')} \, \left[ \int x(t - \tau') \,  x^*(t - \tau) e^{-j 2 \pi (\nu - \nu') (t - \tau)} \, dt \, \right] \, d\tau'  \, d\nu' \,\, + \, \int n(t) \, x^*(t - \tau) \, e^{-j 2 \pi \nu (t - \tau)} \, dt \nonumber \\
    & & \hspace{-21mm} =  \iint h(\tau', \nu') \, e^{j 2 \pi \nu' (\tau - \tau')} \, \underbrace{\left[ \int x(t') \, x^*(t' - (\tau - \tau')) \, e^{-j 2 \pi (\nu - \nu') (t' - (\tau - \tau'))} \, dt' \right]}_{= A_{x,x}(\tau - \tau', \nu - \nu')} d\tau' \, d\nu' \, + \, \int n(t) \, x^*(t - \tau) \, e^{-j 2 \pi \nu (t - \tau)} \, dt \nonumber \\
    & & \hspace{-21mm} =  \underbrace{\iint h(\tau', \nu') \, A_{x,x}(\tau - \tau', \nu - \nu') \, e^{j 2 \pi \nu' (\tau - \tau')}  d\tau' \, d\nu'}_{= \, h(\tau, \nu) \, \ast_{\sigma} \, A_{x,x}(\tau, \nu) } \, + \, \int n(t) \, x^*(t - \tau) \, e^{-j 2 \pi \nu (t - \tau)} \, dt.
\end{eqnarray}
\begin{eqnarray*}
    \hline
\end{eqnarray*}
\end{figure*}

\section{Derivation of Eq.(\ref{ch2_eqninnerproductdd})}
\label{app3}
Using (\ref{zaktransformeqn}), 
the DD domain representations of $a(t)$ and $b(t)$ are given by
\begin{eqnarray}
  a_{_{\mbox{\footnotesize{dd}}}}(\tau , \nu) & = & \sqrt{\tau_p} \sum\limits_{k_1 \in {\mathbb Z}} a(\tau + k_1 \tau_p) \, e^{-j 2 \pi k_1 \nu \tau_p}, \nonumber \\
  b_{_{\mbox{\footnotesize{dd}}}}(\tau , \nu) & = & \sqrt{\tau_p} \sum\limits_{k_2 \in {\mathbb Z}} b(\tau + k_2 \tau_p) \, e^{-j 2 \pi k_2 \nu \tau_p}.
\end{eqnarray}
Using these expressions in the RHS in (\ref{ch2_eqninnerproductdd}) gives (\ref{app3_eqn2}) which completes the derivation.
\begin{figure*}
\vspace{-5mm}
\begin{eqnarray}
\label{app3_eqn2}
  \int\limits_{0}^{\tau_p} \int\limits_{0}^{\nu_p} a_{_{\mbox{\footnotesize{dd}}}}(\tau , \nu) \, b^*_{_{\mbox{\footnotesize{dd}}}}(\tau , \nu) \, d\nu \, d\tau &  =  & \sum\limits_{k_1 \in {\mathbb Z}}  \sum\limits_{k_2 \in {\mathbb Z}} \int\limits_{0}^{\tau_p} \tau_p \int\limits_{0}^{\nu_p} a(\tau + k_1 \tau_p) \, b^*(\tau + k_2 \tau_p) \, e^{j2 \pi \nu (k_2 - k_1) \tau_p} \, d\nu \, d\tau \nonumber \\
  & & \hspace{-40mm} = \sum\limits_{k_1 \in {\mathbb Z}}  \sum\limits_{k_2 \in {\mathbb Z}} \int\limits_{0}^{\tau_p} a(\tau + k_1 \tau_p) \, b^*(\tau + k_2 \tau_p) \, \left[   \tau_p  \int\limits_{0}^{\nu_p} e^{j2 \pi \nu (k_2 - k_1) \tau_p} \, d\nu  \right]  \, d\tau \nonumber \\
  & &  \hspace{-40mm} =   \sum\limits_{k_1 \in {\mathbb Z}}  \sum\limits_{k_2 \in {\mathbb Z}} \int\limits_{0}^{\tau_p} a(\tau + k_1 \tau_p) \, b^*(\tau + k_2 \tau_p) \,  \delta[k_2 - k_1 ] d\tau \nonumber \\
  & & \hspace{-40mm} =   \sum\limits_{k_1 \in {\mathbb Z}}  \int\limits_{0}^{\tau_p} a(\tau + k_1 \tau_p) \, b^*(\tau + k_1 \tau_p) \, d\tau \, = \, \int\limits_{-\infty}^{\infty} a(t) \, b^*(t) \, dt.
\end{eqnarray}
\begin{eqnarray*}
    \hline
\end{eqnarray*}
\end{figure*}

\section{Derivation of Eq.(\ref{iorelation})}
\label{app4}
Let 
\begin{eqnarray}
\label{app4_eqn1}
    {\Tilde y}(t) & \Define & \iint h(\tau, \nu) \, x(t - \tau) \, e^{j 2 \pi \nu (t - \tau) } \, d\tau \, d\nu 
\end{eqnarray} denote the noise-free component of $y(t)$. Then clearly
\begin{eqnarray}
    y_{_{\mbox{\footnotesize{dd}}}}(\tau, \nu) & = & {\mathcal Z}_t\left( y(t)\right) \nonumber \\
    & = & {\Tilde y}_{_{\mbox{\footnotesize{dd}}}}(\tau, \nu) \, + \, {n}_{_{\mbox{\footnotesize{dd}}}}(\tau, \nu)
\end{eqnarray}where ${n}_{_{\mbox{\footnotesize{dd}}}}(\tau, \nu)$ is the Zak-transform of
$n(t)$ and ${\Tilde y}_{_{\mbox{\footnotesize{dd}}}}(\tau, \nu)$ is the Zak-transform of ${\Tilde y}(t)$ which is given by (\ref{iorefprf}) (see top of next page). In (\ref{iorefprf}), step (a) follows from (\ref{zaktransformeqn}). In step (b), note that the term within square brackets is the Zak transform of $x(t)$ evaluated at $(\tau - \tau', \nu - \nu')$. Since $x(t)$ is the inverse time-Zak transform of $x_{_{\mbox{\footnotesize{dd}}}}(\tau, \nu)$, it follows that the term within the square brackets is $x_{_{\mbox{\footnotesize{dd}}}}(\tau - \tau', \nu - \nu')$. The last step follows from the definition of the twisted convolution operation in (\ref{twistconv}).
\begin{figure*}
\vspace{-5mm}
\begin{eqnarray}
\label{iorefprf}
{\Tilde y}_{_{\mbox{\footnotesize{dd}}}}(\tau, \nu) & = & {\mathcal Z}_t\left( {\Tilde y}(t)\right)   \mya  \sqrt{\tau_p} \sum\limits_{k=-\infty}^{\infty}   {\Tilde y}(\tau + k \tau_p) \, e^{-j 2 \pi \nu k \tau_p}  \nonumber \\
&  & \hspace{-20mm} =  \sqrt{\tau_p} \sum\limits_{k=-\infty}^{\infty}  \underbrace{\left[ \iint h(\tau',\nu')
x(\tau + k \tau_p - \tau') e^{j 2 \pi \nu' (\tau + k \tau_p - \tau')} d\tau' d\nu'   \right]}_{= {\Tilde y}(\tau + k \tau_p) \, \, (\mbox{\small{see}} \, (\ref{app4_eqn1}))} \, e^{-j 2 \pi \nu k \tau_p}  \nonumber \\
& & \hspace{-20mm} \myb   \iint h(\tau',\nu') \underbrace{\left[ \sqrt{\tau_p} \sum\limits_{k=-\infty}^{\infty} x(\tau -\tau' + k \tau_p )  e^{-j 2 \pi k \tau_p (\nu - \nu')}\right]}_{= x_{_{\mbox{\tiny{dd}}}}(\tau - \tau', \nu - \nu'), \,\, \mbox{\small{see}} \, (\ref{zaktransformeqn})} 
\, e^{j 2 \pi \nu' (\tau - \tau')} \, d\tau' d\nu'  \nonumber \\
& &  \hspace{-20mm} = \iint h(\tau',\nu')  \, x_{_{\mbox{\footnotesize{dd}}}}(\tau - \tau', \nu - \nu') \, e^{j 2 \pi \nu' (\tau - \tau')} \, d\tau' d\nu'   = h(\tau,\nu) \, *_{\sigma} \,  x_{_{\mbox{\footnotesize{dd}}}}(\tau , \nu).  
\end{eqnarray}
\begin{eqnarray*}
    \hline
\end{eqnarray*}
\end{figure*}

\section{Derivation of Eq. (\ref{crossambigdd})}
\label{app5}
Note that $A_{y,x}(\tau, \nu)$ in (\ref{cross_ambiguity_eqn}) is simply the inner product between $y(t)$ and $x(t - \tau) e^{j 2 \pi \nu (t - \tau)}$. The DD domain realization of $y(t)$ is $y_{_{\mbox{\footnotesize{dd}}}}(\tau , \nu)$ and let that for $x(t - \tau) e^{j 2 \pi \nu (t - \tau)}$ be denoted by $g_{_{\mbox{\footnotesize{dd}}}}(\tau , \nu)$.
Since inner products are conserved in the DD domain (see (\ref{ch2_eqninnerproductdd})), it follows that
\begin{eqnarray}
\label{app5_eqn1}
    A_{y,x}(\tau, \nu) & = & \int\limits_{0}^{\tau_p} \int\limits_{0}^{\nu_p} y_{_{\mbox{\footnotesize{dd}}}}(\tau' , \nu' )  \, g_{_{\mbox{\footnotesize{dd}}}}(\tau' , \nu') \, d\tau' \, d\nu'.
\end{eqnarray}$x(t - \tau) e^{j 2 \pi \nu (t - \tau)}$
is simply the output of a channel with input $x(t)$ and having only one path with delay shift and Doppler shift $\tau$ and $\nu$ respectively. Then the DD representation of $x(t - \tau) e^{j 2 \pi \nu (t - \tau)}$ (i.e., $g_{_{\mbox{\footnotesize{dd}}}}(\cdot , \cdot)$) satisfies
\begin{eqnarray}
\label{app5_eqn2}
    g_{_{\mbox{\footnotesize{dd}}}}(\tau', \nu') & = & x_{_{\mbox{\footnotesize{dd}}}}(\tau' - \tau , \nu' - \nu) \, e^{j 2 \pi \nu (\tau' - \tau)}.
\end{eqnarray}Using (\ref{app5_eqn2}) in (\ref{app5_eqn1}) gives (\ref{crossambigdd}).

\section{Proof Of Lemma \ref{lm1}}
\label{appendix_lm1}
Using (\ref{crossambigdd}), the auto-ambiguity function of the time and bandwidth limited chirp signal $u(t)$ is given by
\begin{eqnarray}
\label{lml_eqn1}
    A_{u,u}(\tau, \nu) &  &  \nonumber \\
    & & \hspace{-25mm} = \int\limits_{0}^{\tau_p} \hspace{-1mm} \int\limits_{0}^{\nu_p} \hspace{-1mm} u_{\mbox{\scriptsize{dd}}}(\tau', \nu') \, u_{\mbox{\scriptsize{dd}}}^*(\tau' - \tau, \nu' - \nu) \, e^{-j 2 \pi \nu (\tau' - \tau)} \, d\tau' \, d\nu'.
\end{eqnarray}From {\ref{chirp filter}) we know that
\begin{eqnarray}
\label{lml_eqn2}
     u_{\mbox{\scriptsize{dd}}}(\tau, \nu) & = & w(\tau, \nu) \ast_{\sigma} c_{\mbox{\scriptsize{dd}}}(\tau, \nu) \nonumber \\
     & & \hspace{-30mm} = \iint w(\tau'' , \nu'') c_{\mbox{\scriptsize{dd}}}(\tau - \tau'', \nu - \nu'') \, e^{j 2 \pi \nu'' (\tau - \tau'')} \, d\tau'' \, d\nu''.
\end{eqnarray}Substituting the expression in the R.H.S. of (\ref{lml_eqn2}) in the R.H.S. of (\ref{lml_eqn1}) and
re-arranging the integrals gives (\ref{ambiguityuueqn}).

\section{Proof Of Theorem \ref{thm1} }
\label{appendix_thm1}
From (\ref{ambiguityuueqn}) the ambiguity function of $u(t)$ is given by
\begin{eqnarray}
\label{4}
A_{u,u}(\tau, \nu) & =  & w(\tau, \nu) *_{\sigma} A_{c,c}(\tau, \nu) *_{\sigma} w_{mf}(\tau, \nu), \nonumber \\
& = & I_1(\tau, \nu) \, *_{\sigma} \, w_{mf}(\tau, \nu)
\end{eqnarray}where $A_{cc}(\tau, \nu)$ is
given by (\ref{ct_ambig}) and
\begin{eqnarray} \label{6}
I_1(\tau, \nu) & \Define & w(\tau, \nu) *_{\sigma} A_{c,c}(\tau, \nu),
\end{eqnarray}
Using (\ref{filter eqn}) and (\ref{ct_ambig}) in (\ref{6}) we get (\ref{7}) (see top of this page).
\begin{figure*}
\vspace{-7mm}
\begin{eqnarray}
\label{7}
I_1 &=&  \int_{\tau'} \int_{\nu'} \left(\frac{2 \alpha B^2}{\pi}\right)^{1/4} e^{-\alpha B^2 \tau'^2} \left(\frac{2 \beta T^2}{\pi}\right)^{1/4} e^{-\beta T^2 \nu'^2} \,  e^{j \pi (\nu-\nu') (\tau-\tau')} \, \delta((\nu-\nu') - a (\tau-\tau')) \, e^{j2\pi\nu'(\tau-\tau')}\, d\nu' \, d\tau' \nonumber \\
 &=&  \left(\frac{2 \alpha B^2}{\pi}\right)^{1/4} 
 \, \left(\frac{2 \beta T^2}{\pi}\right)^{1/4}  \, \int_{\tau'}  e^{-\alpha B^2 \tau'^2} \,  e^{-\beta T^2 (\nu-a(\tau-\tau'))^2}\, e^{-j \pi a(\tau-\tau')^2}  e^{j 2  \pi \nu (\tau-\tau')}  \, d\tau' .
\end{eqnarray}
\vspace{-4mm}
\begin{eqnarray*}
\hline
\end{eqnarray*}
\end{figure*}
Using (\ref{7}) in (\ref{4}) we get 
\begin{eqnarray}
\label{9}
A_{u,u}(\tau, \nu) = I_1(\tau, \nu) *_{\sigma} w_{mf}(\tau, \nu).
\end{eqnarray}Using
\begin{eqnarray}
w_{mf}(\tau, \nu) & \hspace{-3mm} = &  \hspace{-3mm}  \left(\frac{2 \alpha B^2}{\pi}\right)^{1/4} \hspace{-2mm} e^{-\alpha B^2 \tau^2} \left(\frac{2 \beta T^2}{\pi}\right)^{1/4} \hspace{-2mm}  e^{-\beta T^2 \nu^2} e^{j2\pi \tau \nu} \nonumber \\
\end{eqnarray}
in (\ref{9}) we get (\ref{10}).
\begin{figure*}
\begin{eqnarray}
\label{10}
\hspace{-30mm} A_{u,u}(\tau, \nu) &=&  \left(\frac{2 \alpha B^2}{\pi}\right)^{1/2} 
 \, \left(\frac{2 \beta T^2}{\pi}\right)^{1/2}  \, \int_{\tau''} \int_{\nu''} \int_{\tau'}  e^{-\alpha B^2 \tau'^2} \,  e^{-\beta T^2 (\nu''-a(\tau''-\tau'))^2}\, e^{-j \pi a(\tau''-\tau')^2}  e^{j 2  \pi \nu'' (\tau''-\tau')}  \, d\tau' \nonumber \\
&& \hspace{50mm} e^{-\alpha B^2 (\tau-\tau'')^2}  e^{-\beta T^2 (\nu-\nu'')^2} e^{j2\pi (\tau-\tau'') (\nu-\nu'')} e^{j 2\pi \nu''(\tau-\tau'')} d\nu'' d\tau'' \nonumber \\
 &=&  \left(\frac{2 \alpha B^2}{\pi}\right)^{1/2} 
 \, \left(\frac{2 \beta T^2}{\pi}\right)^{1/2}  \, \int_{\tau''}  \int_{\tau'}  e^{-\alpha B^2 \tau'^2} \,  e^{-\alpha B^2 (\tau-\tau'')^2} \, \, e^{-j \pi a(\tau''-\tau')^2}  e^{j 2  \pi \nu (\tau-\tau'')} \nonumber \\
&& \hspace{40mm}\underbrace{\int_{\nu''} e^{-\beta T^2 (\nu''-a(\tau''-\tau'))^2} e^{-\beta T^2 (\nu''-\nu)^2}  e^{j2\pi \nu''(\tau''-\tau')} d\nu''}_{I_2}\,\,\, d\tau' d\tau''.
\end{eqnarray}
\vspace{-3mm}
\begin{eqnarray*}
    \hline
\end{eqnarray*}
\end{figure*}
The integral $I_2$ in (\ref{10}) is given by
(\ref{13}), where in the second step we have used the fact that
for any real $c_1, c_2$
\begin{eqnarray}
    \label{14}
\int e^{-c_1(x-c_2)^2}\,\, e^{j2\pi f x}\,\, dx &=& \sqrt{\frac{\pi}{c_1}}\,\,e^{-\frac{\pi^2 f^2}{c_1^2}} \,\,e^{j2\pi c_2 f}.
\end{eqnarray}
\begin{figure*}
\vspace{-4mm}
\begin{eqnarray}
    \label{13}
I_2 &=& e^{-\frac{(a(\tau''-\tau')-\nu)^2}{2}}\int_{\nu''} e^{-2\beta T^2 \left(\nu''-\frac{(\nu+a(\tau''-\tau'))}{2}\right)^2}  e^{j2\pi \nu''(\tau''-\tau')} d\nu'' \nonumber \\
&=& e^{-\frac{(a(\tau''-\tau')-\nu)^2}{2}} \,\,\sqrt{\frac{\pi}{2\beta T^2}} \,\,\, e^{-\frac{\pi^2 (\tau''-\tau')^2}{(2\beta T^2)^2}} \,\, e^{-j \pi (\nu+a(\tau''-\tau'))(\tau''-\tau')}
\end{eqnarray}
\vspace{-3mm}
\begin{eqnarray*}
    \hline
\end{eqnarray*}
\end{figure*}
By using the expression of $I_2$ in (\ref{13}) in the R.H.S. of (\ref{10}) we get (\ref{16}) (see top of next page).
In the second step of (\ref{16}) we have substituted the integration variable $\tau''$ with $\Tilde{\tau} = \tau''-\tau'$.
\begin{figure*}
\vspace{-9mm}
\begin{eqnarray}
\label{16}
\hspace{-30mm} A_{u,u}(\tau, \nu) &=&  \left(\frac{2 \alpha B^2}{\pi}\right)^{1/2} 
 \, \left(\frac{2 \beta T^2}{\pi}\right)^{1/2} \, \sqrt{\frac{\pi}{2\beta T^2}} \, \int_{\tau''}  \int_{\tau'}  e^{-\alpha B^2 \tau'^2} \,  e^{-\alpha B^2 (\tau-\tau'')^2} \, \, e^{-j 2\pi a(\tau''-\tau')^2}  e^{j 2  \pi \nu (\tau-\tau'')}   \nonumber \\
&& \hspace{60mm}\,\,e^{-\frac{(a(\tau''-\tau')-\nu)^2}{2}}\,e^{-\frac{\pi^2 (\tau''-\tau')^2}{(2\beta T^2)^2}} \,\, e^{j \pi \nu (\tau''- \tau')} d\tau' d\tau'' \nonumber \\
&=&  \left(\frac{2 \alpha B^2}{\pi}\right)^{1/2} 
 \, \left(\frac{2 \beta T^2}{\pi}\right)^{1/2} \, \sqrt{\frac{\pi}{2\beta T^2}} \, \int_{\Tilde{\tau}}    e^{-\frac{(a\Tilde{\tau}-\nu)^2}{2}}\,e^{-\frac{\pi^2 \Tilde{\tau}^2}{(2\beta T^2)^2}} e^{-j 2\pi a \Tilde{\tau}^2} \,\, e^{j \pi \nu \Tilde{\tau}} \, e^{j 2  \pi \nu (\tau-\Tilde{\tau})}   \nonumber \\
&& \hspace{60mm} \underbrace{\int_{\tau'} e^{-\alpha B^2 \tau'^2} \,  e^{-\alpha B^2 (\tau-\tau'-\Tilde{\tau})^2} \, e^{-j2\pi \nu \tau'}\, \,  d\tau'}_{I_3} \,\,\, d\Tilde{\tau}.
\end{eqnarray}
\vspace{-3mm}
\begin{eqnarray*}
    \hline
\end{eqnarray*}
\end{figure*}
The integral $I_3$ in (\ref{16}) is given by
\begin{eqnarray}
    \label{18}
    I_3 &=& \int_{\tau'} e^{-2\alpha B^2 \left( \tau'^2 +\frac{(\tau-\Tilde{\tau})^2}{2} -\tau'(\tau-\Tilde{\tau}) \right)} \, e^{-j2\pi \nu \tau'}\, \,  d\tau' \nonumber \\
    & &  \hspace{-8mm} = e^{-\frac{\alpha B^2(\tau-\Tilde{\tau})^2}{2}}  \int_{\tau'} e^{-2\alpha B^2 \left( \tau' -\frac{(\tau-\Tilde{\tau})}{2}  \right)^2} \, e^{-j2\pi \nu \tau'}\,  d\tau'.
\end{eqnarray}Substituting integration variable $\tau'$ with
$t = \tau' - \frac{(\tau - \Tilde{\tau})}{2}$, we get
\begin{eqnarray}
    \label{19}
    I_3 &=& e^{-\frac{\alpha B^2(\tau-\Tilde{\tau})^2}{2}} \,\, e^{-j\pi \nu (\tau - \Tilde{\tau})}\int_{\tau'} e^{-2\alpha B^2 t^2} \, e^{-j2\pi \nu t}\, dt \nonumber \\
\end{eqnarray}
By using (\ref{14}) in (\ref{19}) we get 
\begin{eqnarray}
    \label{20}
    I_3 &=& e^{-\frac{\alpha B^2(\tau-\Tilde{\tau})^2}{2}} \,\, e^{-j\pi \nu (\tau - \Tilde{\tau})}  \sqrt{\frac{\pi}{2\alpha B^2}} \,\, e^{-\frac{\pi^2 \nu^2}{(2\alpha B^2)^2}} .
\end{eqnarray}Substituting (\ref{20}) in (\ref{16}) we finally get
the expression of $A_{u,u}(\tau, \nu)$ in (\ref{Auufinaleqn}).

\section{Derivation of the auto-ambiguity function of Zak-OTFS pulsone}
\label{appn6}
From the DD domain expression for the ambiguity function
in (\ref{crossambigdd}) it follows that the auto-ambiguity
of the Zak-OTFS pulsone is given by

{\vspace{-4mm}
\small
\begin{eqnarray}
      A_{p,p}(\tau, \nu) &  =  & \int\limits_{0}^{\tau_p} \hspace{-1mm} \int\limits_{0}^{\nu_p} \hspace{-1mm} {\Big [} p_{_{\mbox{\scriptsize{dd}},0,0}}^w(\tau', \nu') \, {p_{_{\mbox{\scriptsize{dd}},0,0}}^{w^*}}(\tau' - \tau, \nu' - \nu) \nonumber \\
      & & \hspace{10mm} e^{-j 2 \pi \nu (\tau' - \tau)} \, {\Big ]} d\tau' \, d\nu'.
\end{eqnarray}\normalsize}
Since $p_{_{\mbox{\scriptsize{dd}},0,0}}^w(\tau, \nu) = w(\tau, \nu) \, \ast_{\sigma} \, p_{_{\mbox{\scriptsize{dd}},0,0}}(\tau, \nu)$, from (\ref{ambiguityuueqn}) of Lemma \ref{lm1} it follows that
\begin{eqnarray}
    A_{p,p}(\tau, \nu) &  =  & w(\tau, \nu) \, \ast_{\sigma} \, {\Big (}  \sum\limits_{n \in {\mathbb Z}} \sum\limits_{m \in {\mathbb Z}} \delta(\tau - n \tau_p) \, \delta(\nu - m \nu_p) {\Big )} \nonumber \\
    & & \hspace{6mm} \, \ast_{\sigma} \, w_{\mbox{\tiny{mf}}}(\tau, \nu)
\end{eqnarray}since the auto-ambiguity function of $p_{_{\mbox{\scriptsize{dd}},0,0}}(\tau, \nu)$ is ${\Big (}  \sum\limits_{n \in {\mathbb Z}} \sum\limits_{m \in {\mathbb Z}} \delta(\tau - n \tau_p) \, \delta(\nu - m \nu_p) {\Big )}$ (as we shall show next). The auto-ambiguity of $p_{_{\mbox{\scriptsize{dd}},0,0}}(\tau, \nu)$ is the same as that for its TD realization $p_{0,0}(t) = \sqrt{\tau_p} \sum\limits_{n \in {\mathbb Z}} \delta(t - n \tau_p)$
which is given by (\ref{eqnpt}). Note that in step (a), we substitute the inner summation index $n_1$ by $n = n_1 - n_2$
to arrive at step (b). The last step follows from the fact that $\tau_p \sum\limits_{n_2 \in {\mathbb Z}} e^{-j 2 \pi n_2 \tau_p}  = \sum\limits_{m \in {\mathbb Z}} \delta(\nu - \frac{m}{\tau_p})$ and $\nu_p = 1/\tau_p$.
\begin{figure*}
\vspace{-9mm}
\begin{eqnarray}
\label{eqnpt}
\int p_{0,0}(t) \, p_{0,0}^*(t - \tau) \, e^{-j 2 \pi \nu (t - \tau)} \, dt & = & \tau_p \sum\limits_{n_1, n_2 \in {\mathbb Z}} \int \delta(t - n_1 \tau_p) \, \delta(t - \tau - n_2 \tau_p) \, e^{-j 2 \pi \nu (t - \tau)} \, dt \nonumber \\
& \mya & \tau_p \sum\limits_{n_2 \in {\mathbb Z}} {\Big (} \sum\limits_{n_1 \in {\mathbb Z}} \delta(\tau - (n_1 - n_2) \tau_p {\Big )} \, e^{-j 2 \pi n_2 \tau_p}  \nonumber \\
& \myb & \tau_p \sum\limits_{n_2 \in {\mathbb Z}} {\Big (} \sum\limits_{n \in {\mathbb Z}} \delta(\tau - n \tau_p {\Big )} \, e^{-j 2 \pi n_2 \tau_p} \nonumber \\
& = & \sum\limits_{n \in {\mathbb Z}} \sum\limits_{m \in {\mathbb Z}} \delta(\tau - n \tau_p) \delta(\nu - m \nu_p)
\end{eqnarray}
\vspace{-3mm}
\begin{eqnarray*}
    \hline
\end{eqnarray*}
\vspace{-5mm}
\end{figure*}

\end{document}